\documentclass{emulateapj}
\usepackage{natbib,times}
\usepackage{setspace}

\newcommand{\mc}{\multicolumn}

\shorttitle{Calibration of the optical mass proxy for clusters of galaxies}
\shortauthors{Z. L. Wen \& J. L. Han}

\begin{document}

\title{Calibration of the optical mass proxy for clusters of galaxies and 
an update of the\\ WHL12 cluster catalog}

\author{Z. L. Wen, \and J. L. Han}
\affil{National Astronomical Observatories, 
                Chinese Academy of Sciences, 
                20A Datun Road, Chaoyang District, Beijing 100012, China; 
                zhonglue@nao.cas.cn.}


\begin{abstract}

Accurately determining the mass of galaxy clusters is fundamental for
many studies on cosmology and galaxy evolution. We collect and rescale
the cluster masses of 1191 clusters of $0.05<z<0.75$ estimated by
X-ray or Sunyaev--Zeldovich measurements and use them to calibrate the
optical mass proxy. The total $r$-band luminosity (in units of
$L^{\ast}$) of these clusters are obtained by using spectroscopic and
photometric data of the Sloan Digital Sky Survey (SDSS). We find that
the correlation between the cluster mass $M_{500}$ and total $r$-band
luminosity $L_{500}$ significantly evolves with redshift. After
correcting for the evolution, we define a new cluster richness
$R_{L\ast,500}=L_{500}\,E(z)^{1.40}$ as the optical mass proxy. By
using this newly defined richness and the recently released SDSS DR12
spectroscopic data, we update the WHL12 cluster catalog and identify
25,419 new rich clusters at high redshift. In the SDSS spectroscopic
survey region, about 89\% of galaxy clusters have spectroscopic
redshifts. The mass can be estimated with a scatter of 0.17 dex for
the clusters in the updated catalog.

\end{abstract}

\keywords{galaxies: clusters: general --- galaxies: distances and redshifts}

\section{Introduction}

Clusters of galaxies are the most massive gravitationally bound systems
in the universe. Statistics of cluster properties provides very
powerful constraints on cosmology and galaxy evolution
\citep[see][]{aem11,wtc12}. The cluster mass is a fundamental
parameter for galaxy clusters and is related to observational
features in the X-ray, optical, millimeter and radio bands
\citep[e.g.,][]{voi05,bcd+09}. Galaxies, intracluster hot gas and dark
matter contribute about 5\%, 15\%, and 80\% of the total cluster mass,
respectively. Determining the cluster mass is the basis for many
studies. For example, the mass function for a complete sample of
galaxy clusters can be used to constrain cosmological parameters
\citep[e.g.,][]{rb02,vkb+09,whl10}.

Statistics for weak gravitational lensing features is the most direct
method to determine the total cluster mass including contributions
from member galaxies, hot intracluster gas and dark matter. However,
this method requires high-quality image data and also redshift
information for faint background galaxies and thus only a few massive
clusters have their masses so determined
\citep[e.g.,][]{dah06,bsk+07,otu+10,oaa+14}. Based on the assumption that the
hot intracluster gas is in hydrostatic equilibrium, the cluster
mass can be estimated by using the X-ray surface brightness and
temperature distributions \citep[e.g., ][]{frb01,rb02,kvn06,vkf+06}. A
few thousands of galaxy clusters have their masses estimated from the
X-ray observables \citep{vbe+09,mae+10,pap+11,tsl13}. In the
millimeter band, the thermal Sunyaev--Zeldovich (SZ) effect provides 
robust mass estimates for more than one thousand massive clusters
\citep{hhm+13,rsb+13,planck14}, which is independent of cluster
dynamical state and redshift \citep{che02}.

The velocity dispersion of member galaxies is certainly a good measure
of cluster mass \citep{ggm+98}, which can be derived from optical
spectroscopic data and have been obtained for a limited number of
galaxy clusters. However, the spectroscopic observations are usually
incomplete for cluster member galaxies, and cluster substructures
often induce a bias on the estimated mass
\citep{bir95,smh+13}.
On the other hand, photometric data can be used to identify galaxy
clusters
\citep[e.g.,][]{kma+07,whl09,hmk+10,spd+11,whl12,ogu14,rrb+14}. Cluster
richness, defined as the total number of member galaxies or their
total luminosity, can be used as an optical mass proxy
\citep{pbb+05,pbb+07}. The challenge is to accurately determine the
membership of cluster galaxies by eliminating the contamination by
field galaxies. Based on multicolor survey data, the member galaxies
can be discriminated by using galaxy colors
\citep[e.g.,][]{gy00,kma+07,hmk+10} or photometric redshifts
\citep[e.g.,][]{whl09,spd+11}. The cluster richness derived from
optical photometric data in general is poorly correlated with cluster
mass. However, if the correlation is improved, the cluster richness
can be used to estimate cluster mass for a very large sample of
clusters even up to high redshifts.

The Sloan Digital Sky Survey \citep[SDSS;][]{yaa+00} offers an
unprecedented photometric data in five broad bands ($u$, $g$, $r$, $i$,
and $z$) covering 14,000 deg$^2$ with the exceptional follow-up
spectroscopic observations. The photometric data reach a limit of
$r=22.2$ \citep{slb+02}, with the star--galaxy separation reliable to a
limit of $r=21.5$ \citep{lgi+01}. The spectroscopic survey observes
galaxies with a Petrosian magnitude of $r<17.77$ for the main galaxy
sample \citep{swl+02} and a model magnitude $r<19.9$ for the Luminous
Red Galaxy sample \citep{eag+01}. These galaxies cover a footprint of
10,400 deg$^2$ in the range of $-11^{\circ}\leq {\rm Decl.}
\leq 69^{\circ}$. The latest Data Release 12 \citep[DR12,][]{dr12+15}
includes the photometric data for about 200 million galaxies and the
spectroscopic data for about 2.32 million galaxies.

The large samples of galaxy clusters or groups have been identified
by using the SDSS spectroscopic data
\citep[e.g.,][]{mz05,bfw+06,ttg+14} and the photometric data
\citep[e.g.,][]{kma+07,whl09,hmk+10,spd+11,whl12,ogu14,rrb+14}. Generally,
the richness in these cluster catalogs is defined as the total number
or the total luminosity of member galaxies, which can be regarded as a
coarse mass proxy for clusters up to a redshift of $z\sim0.4$. At
higher redshifts, the discrimination of member galaxies are limited to
only very luminous galaxies, and hence the richness is systematically
underestimated, as discussed in \citet{whl12}. We noticed that without
a proper calibration, such a cluster richness could be
redshift-dependent even at low redshift \citep[e.g.,][]{rrk+09,whl10}.

Many recent samples of galaxy clusters have their masses determined by
X-ray or SZ measurements up to redshift of $z\sim0.75$
\citep[e.g.,][]{pap+11,planck14}. These provide an opportunity to
calibrate the richness over a large redshift range so that the
calibrated richness may be used as a mass proxy for a very large
sample of galaxy clusters. In Section 2, we collect the cluster mass
in the literature estimated by the X-ray or SZ observations, and for
these clusters in the SDSS survey region, we calculate their total
$r$-band luminosity and define a mass proxy that is independent of
redshift. On the other hand, the SDSS Baryon Oscillation Spectroscopic
Survey \citep{dsa+13} offers spectroscopic redshifts for most of the
galaxy clusters in the catalog of \citet[][i.e., the WHL12 cluster
  catalog]{whl12}, so that in Section 3 we update the richness and
redshift information for the WHL12 clusters, and find new
complementary clusters at high redshift.

Throughout this paper, we assume a $\Lambda$CDM cosmology, taking
$H_0=100 h$ km~s$^{-1}$ Mpc$^{-1}$, with $h=0.7$, $\Omega_m=0.3$
and $\Omega_{\Lambda}=0.7$.


\section{Calibrating the richness for an optical mass proxy}

We collect the estimated masses of many large samples of galaxy
clusters, and then scale them to each other by using common clusters.
The scaled masses of the composite sample of galaxy clusters are used
to calibrate the optical richness of galaxy clusters in a large range
of redshifts so that any redshift dependence can be eliminated.

\begin{figure*}
  \centering
  \resizebox{0.3\textwidth}{!}{\includegraphics{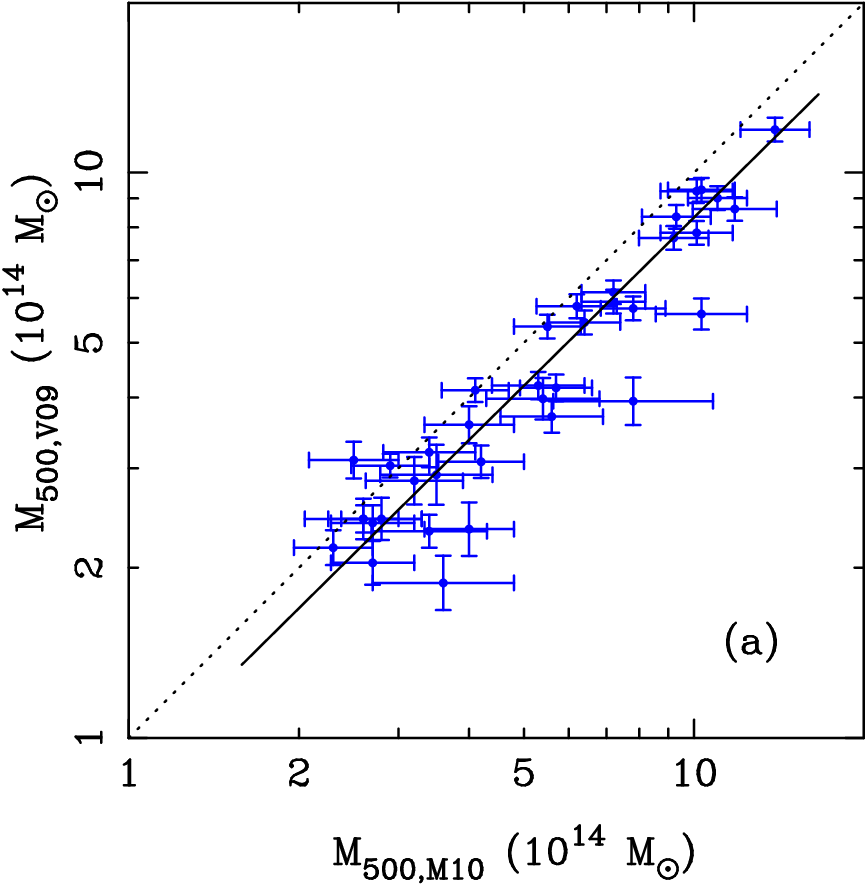}} \hspace{3mm}
  \resizebox{0.3\textwidth}{!}{\includegraphics{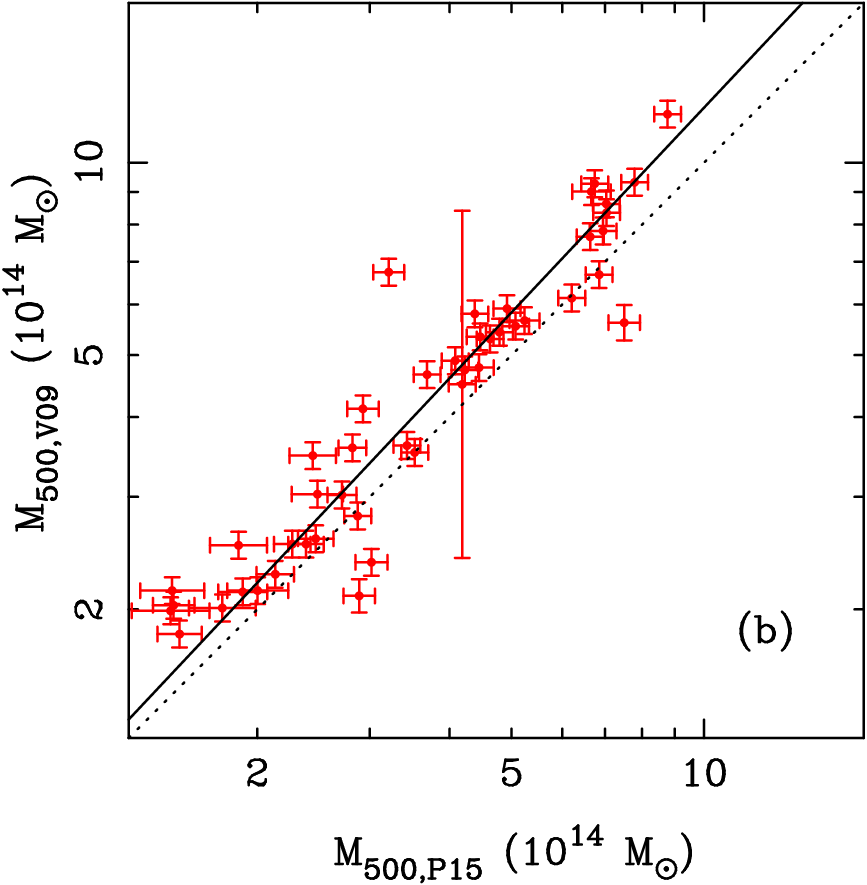}} \hspace{3mm}
  \resizebox{0.3\textwidth}{!}{\includegraphics{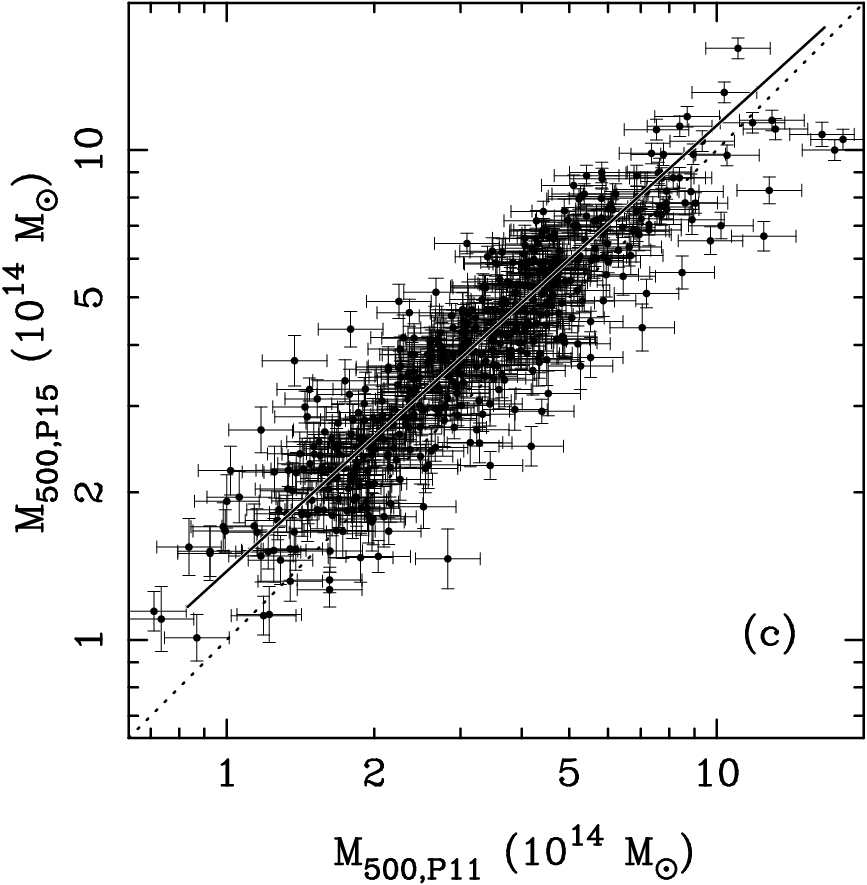}}\\[3mm]
  \resizebox{0.3\textwidth}{!}{\includegraphics{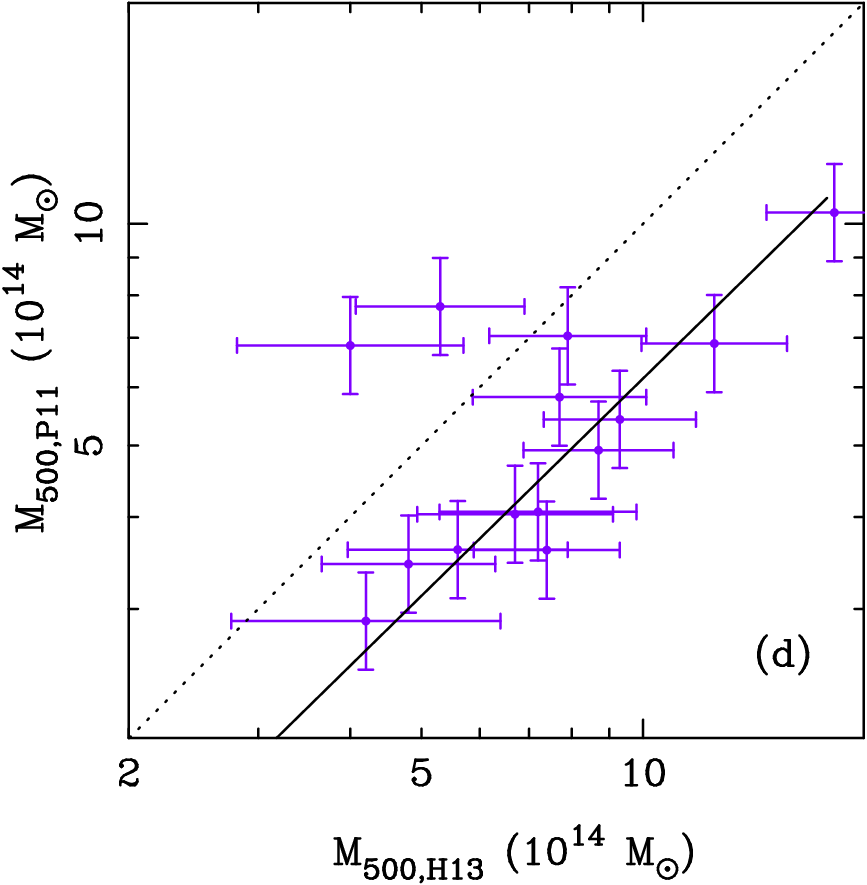}} \hspace{3mm}
  \resizebox{0.3\textwidth}{!}{\includegraphics{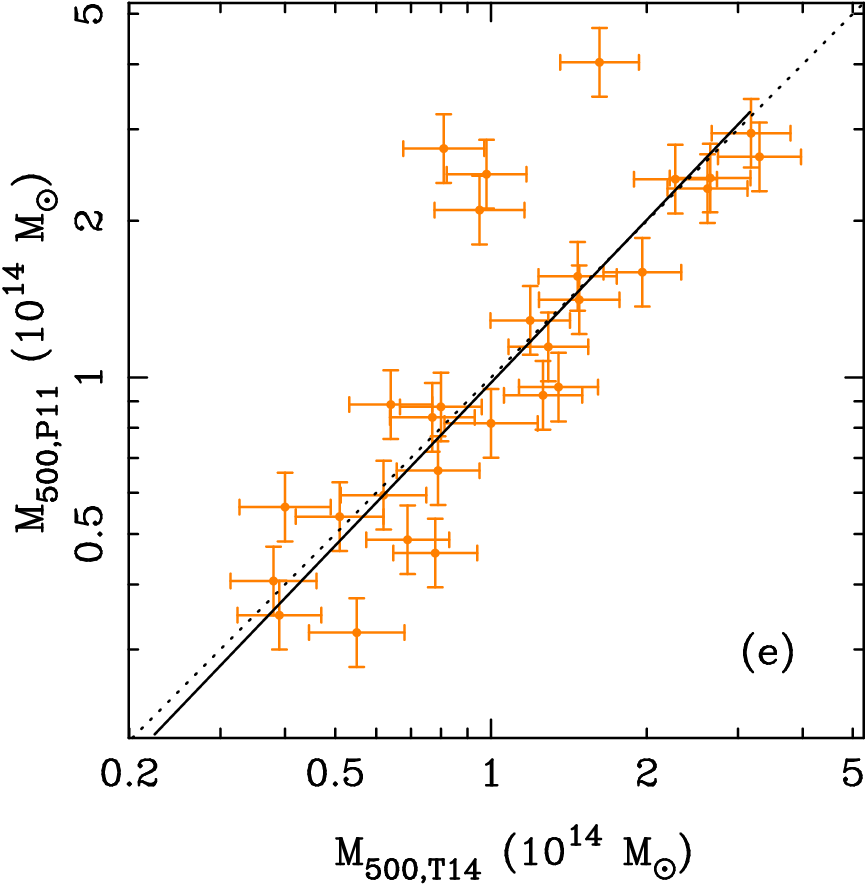}} \hspace{3mm}
  \resizebox{0.3\textwidth}{!}{\includegraphics{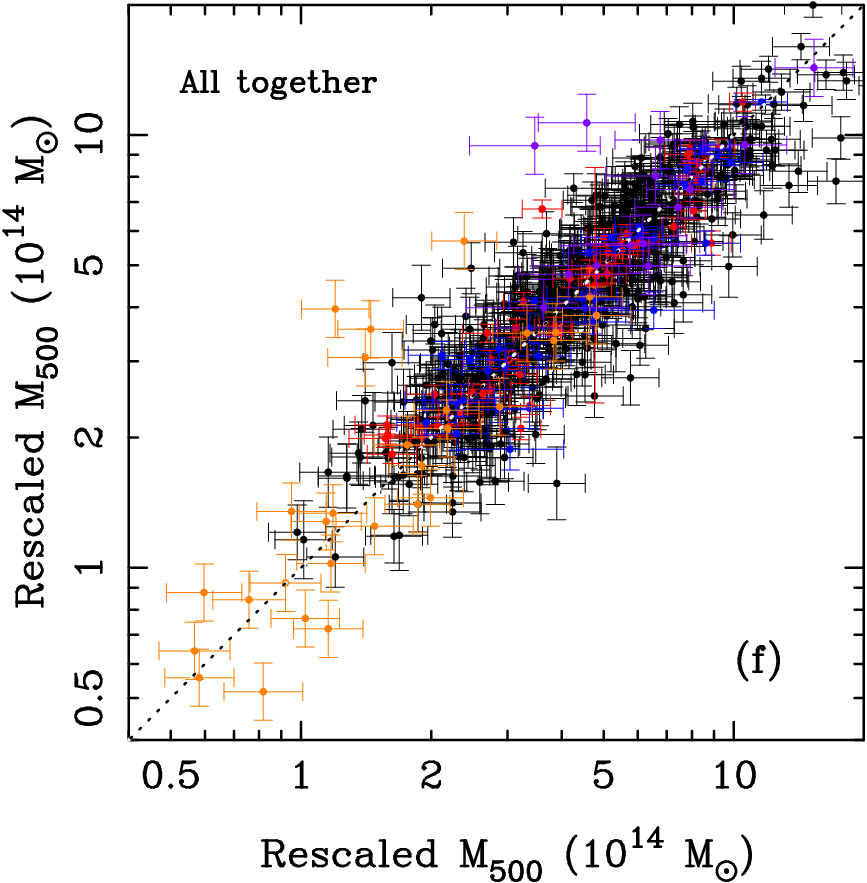}}
  \caption{Comparison of the estimated masses for
   the overlapping common clusters in different samples from the literature. The
    solid line is the best fit, and the dotted line indicates the
    equivalence. All data in panels (a) -- (e) are finally rescaled to
    V09 via the best fit(s), and summarized in panel (f), where
    the mass values in the X-axis are rescaled from the X-values of
    panels (a) -- (e) and the mass values in the Y-axis are rescaled 
    from the Y-values of panels (a) -- (e). Because the
    intrinsic uncertainty of mass-observable relation is larger than 5\%
    \citep{vbe+09}, we here set the minimum uncertainty of 5\% for
    cluster mass even though it could be formally small in the literature.
\label{norma}}
\end{figure*}

\subsection{The mass estimates for galaxy clusters and their
  rescaling between different samples}

Many samples of galaxy clusters have their mass estimated by using
the X-ray or SZ measurements
\citep[e.g.][]{kvn06,zfb+08,pca+09,zjc+13}. To make the largest
composite sample with unified scaling on mass estimates, however,
we collect only large samples of clusters from the literature, with each
having at least 10 overlapping common clusters with the other samples so
that the estimated masses of these common clusters can be compared
and scaled. For this purpose, we take six large samples from
the literature: the X-ray cluster samples by \citet[][ hereafter
  V09]{vbe+09}, \citet[][hereafter M10]{mae+10}, \citet[][ hereafter
  P11]{pap+11} and \citet[][hereafter T14]{tsl13,tsl14}, and the SZ
cluster samples from the Atacama Cosmology Telescope (ACT) survey by
\citet[][hereafter H13]{hhm+13} and the Planck survey by the 
\citet[][hereafter P15]{planck15}. We extract the cluster mass
$M_{500}$ from these references, which is the mass within a radius of
$r_{500}$ within which the mean density of a cluster is 500 times of
the critical density of the universe, $\rho_c\equiv 3H^2(z)/8\pi G$,
i.e.
\begin{equation}
M_{500}=\frac{4\pi}{3}r^3_{500}\times 500\rho_c.
\label{m500def}
\end{equation}

In the following we briefly introduce the six samples:\\[2mm]
\noindent {\bf V09:} \citet{vbe+09} estimated the total masses for 85
clusters at $z\approx0.05$ and $z\approx0.5$ by using high-quality
{\it Chandra} observations. The cluster mass $M_{500}$ was derived by
using three X-ray proxies: average temperature $T_X$, gas mass $M_{\rm
  gas}$, and $Y_X=T_X\times M_{\rm gas}$. \citet{vbe+09} calibrated
their scaling relations with cluster mass \citep{kvn06} by using a
sample of well--observed relaxed clusters at low redshifts. The mass
estimates derived using the three proxies are well matched with very
small scatters. We adopt the cluster mass $M_{500}$ they derived from
$M_{\rm gas}$ and the gas mass fraction, which depends weakly on
cluster mass.  \\[2mm]
\noindent
{\bf M10:} \citet{mae+10} used high-quality {\it Chandra} observations
to estimate the gas mass $M_{\rm gas}$ for 130 massive clusters that
have 0.1-2.4 keV luminosity larger than
$2.5\times10^{44}$erg~s$^{-1}$ in the redshift range of
$0.05<z<0.88$. Different from V09, they derived the total mass
$M_{500}$ assuming a constant value of gas mass fraction.  \\[2mm]
\noindent
{\bf P11:} \citet{pap+11} compiled 1743 clusters in the redshift range
of $z<1.2$ from the ROSAT All Sky Survey-based data, including
NORAS \citep{bvh+00}, REFLEX \citep{bsg+04}, BCS \citep{eeb+98}, SGP
\citep{cvb+02}, NEP \citep{hmv+06}, MACS \citep{eeh01} and CIZA
\citep{emt02}, and serendipitous searches in the ROSAT observations
including 160SD \citep{mmq+03}, 400SD \citep{bvh+07}, SHARC
\citep{rnh+00}, WARPS \citep{hpe+08} and EMSS \citep{gms+90}. All
clusters have their X-ray luminosity, $L_{X,500}$, measured within
$r_{500}$. The cluster masses were estimated from $L_{X,500}$ by
using the scaling relation between mass and X-ray luminosity obtained
by \citet{app+10} and \citet{pca+09}, with a typical uncertainty of
15\%.  \\[2mm]
\noindent {\bf T14:} \citet{tsl13} presented a sample of 530 X-ray
clusters selected from the XMM-Newton serendipitous source catalog
in the SDSS sky region, of which 310 clusters have spectroscopic
redshifts and are in the redshift range of
$0.03<z<0.70$. \citet{tsl14} updated the photometric redshifts of the
clusters by using the SDSS DR10 spectroscopic data. Cluster masses
are estimated by using the X-ray luminosity $L_{X,500}$ and the
scaling relations given by \citet{pca+09}.
\\[2mm]
\noindent {\bf H13:} \citet{hhm+13} presented a catalog of 68 clusters
over $0.15<z<0.8$ detected via SZ effect at 148 GHz in the ACT
survey. The authors provided cluster mass estimates for the 68
clusters near the celestial equator together with 23 clusters in the
southern survey. Four different mass estimates were given in the paper
based on the Compton parameter calibrated to the Universal Pressure
Profile model \citep{app+10}, the \citet{boc+12} model, the Nonthermal
model \citep{tbo11} and the dynamical masses of \citet{smh+13},
respectively. These four mass estimates are well correlated with
different normalization. Here, we adopt the mass estimates calibrated
to dynamical masses of \citet{smh+13}.
\\[2mm]
\noindent {\bf P15:} The all-sky Planck catalog of clusters is the
largest SZ-selected catalog to date, containing 1653 clusters with
redshifts up to $z\sim1$ in the second catalog \citep{planck15}. Among
these, 1203 clusters have counterparts in previous X-ray, optical, SZ, 
and infrared data and 1094 clusters have redshifts. The masses of the
1094 clusters with redshift information were derived by using the
Compton parameter \citep{planck14}.

The masses of galaxy clusters in these samples are compiled to form a
very large composite sample. However, we notice that cluster masses in
these six samples were estimated by using different mass proxies with
different scaling relations, and even cluster masses derived from
different X-ray proxies in the same sample are not consistent as
well. Though both V09 and M10 derived cluster masses via the gas mass
$M_{\rm gas}$, normalization of the fraction of gas mass and
dependence on cluster mass are different in the two papers. Different
cluster masses are obtained in H13 if they are normalized to different
models. Therefore, to get the uniformed scale for estimated masses of
galaxy clusters in a composite sample, the mass estimates in different
samples by different authors must be rescaled to each other for
consistency.

\begin{table*}
{\footnotesize
\begin{center}
  \caption{Rescaled mass and newly calculated richness
    for 1191 galaxy clusters}
\label{tab1}
\begin{tabular}{rrrcccccrc}
\hline
\mc{1}{l}{ID} &\mc{1}{c}{R.A.}  &\mc{1}{c}{Decl.}  &\mc{1}{c}{$z$}  & \mc{1}{c}{$M_{500}$} & 
\mc{1}{c}{$\sigma_{M_{500}}$} & \mc{1}{c}{$r_{500}$} & \mc{1}{c}{$\sigma_{r_{500}}$} & 
$R_{L\ast,500}$ & \mc{1}{l}{Reference}\\
              &\mc{1}{c}{(deg)} &\mc{1}{c}{(deg)} &                &  \mc{1}{c}{($10^{14}~M_{\odot}$)} &
\mc{1}{c}{($10^{14}~M_{\odot}$)} & (Mpc)  & (Mpc)  &           &    \\
(1) & \mc{1}{c}{(2)} & \mc{1}{c}{(3)} & (4) & (5) & (6) & (7) & (8) & \mc{1}{c}{(9)} & (10)\\
\hline   
   1 &  0.49367 & 12.06612 & 0.1995 &  4.40 &  0.42 & 1.08 & 0.04&  86.20 & P11,P15\\             
   2 &  0.79826 &$-6.09169$& 0.2328 &  7.20 &  0.44 & 1.26 & 0.03& 105.02 & P11,P15\\             
   3 &  0.95698 &  2.06647 & 0.0967 &  2.57 &  0.26 & 0.94 & 0.03&  60.13 & P11,P15\\             
   4 &  1.24353 & 11.70091 & 0.0763 &  2.00 &  0.30 & 0.87 & 0.05&  23.18 & P11    \\             
   5 &  1.34984 & 16.21926 & 0.1155 &  3.70 &  0.56 & 1.05 & 0.06&  67.62 & P11    \\             
   6 &  1.58453 & 10.86429 & 0.1680 &  4.46 &  0.41 & 1.10 & 0.03&  68.22 & P11,P15\\             
   7 &  2.04332 &  2.02009 & 0.3665 &  6.44 &  0.70 & 1.16 & 0.04&  68.70 & H13,P15\\             
   8 &  2.33609 &  6.82309 & 0.2378 &  5.80 &  0.66 & 1.17 & 0.05& 112.20 & P15    \\             
   9 &  2.51443 & 17.77082 & 0.1714 &  3.76 &  0.59 & 1.04 & 0.06&  64.06 & P15    \\             
  10 &  2.57239 &  6.64203 & 0.2648 &  5.68 &  0.72 & 1.15 & 0.05&  53.00 & P15    \\ 
\hline
\end{tabular}
\end{center}
{Notes:
Column (1): sequence number;
Column (2) and (3): R.A. (J2000) and Decl. (J2000) of cluster;
Column (4): spectroscopic redshift of cluster;
Column (5) and (6): rescaled cluster mass $M_{500}$ and its uncertainty $\sigma_{M_{500}}$;
Column (7) and (8): recalculated cluster radius $r_{500}$ and its uncertainty $\sigma_{r_{500}}$ from $M_{500}$;
Column (9): newly defined cluster richness $R_{L\ast,500}$;
Column (10): reference for original mass estimate: V09 for \citet{vbe+09}, M10 for \citet{mae+10}, 
P11 for \citet{pap+11}, H13 for \citet{hhm+13}, P15 for \citet{planck15}, T14 for \citet{tsl13,tsl14}.\\
{\small (This table is available in its entirety in a machine-readable form in
the online journal. A portion is shown here for guidance regarding the
form and content.)}
}
}
\end{table*} 

In general, the mass estimates derived by the proxies of $M_{\rm gas}$
and $Y_X$ are more reliable than those from the X-ray luminosity
\citep{kvn06}. For the SZ clusters, we find that the mass estimates
depend on model for the ACT clusters \citep{hhm+13}; the cluster
masses from the Planck survey are probably systematically underestimated
\citep{planck14,vma+14}. In this work, we rescale cluster masses to the
standard obtained by \citet{vbe+09} because their mass estimates are
based on very high-quality X-ray data and the mass proxy shows only
a small scatter after careful calibration using the known masses of low
redshift relaxed clusters.

The easy approach to rescale the cluster masses to the standard of
\citet{vbe+09} is to compare the mass values of common clusters
that overlap in these different samples. As shown in
Figure~\ref{norma}(a), cluster masses derived by V09 and M10 using the
same proxy $M_{\rm gas}$ show a systematic offset. The best fit
is found as
\begin{equation}
\log M_{\rm 500,V09}=(0.99\pm0.06)\log M_{\rm 500,M10}-(0.07\pm0.04).
\end{equation}
This equation can be used to convert the mass estimates of all galaxy
clusters in the M10 sample to the V09 standard. Similarly, we compare
the mass estimates for overlapping common galaxy clusters in V09 and
P15, as shown in Figure~\ref{norma}(b), and obtain the conversion equation
as
\begin{equation}
\log M_{\rm 500,V09}=(1.06\pm0.03)\log M_{\rm 500,P15}+(0.02\pm0.02).
\label{v09p13}
\end{equation}

In the cluster sample of P11, there are too few clusters that overlap
with V09 sample. However, there are many overlapping common clusters
with the P15 catalog. We can rescale the cluster masses of the P11
sample to the P15 scale first, and then convert to the V09 standard.
Comparing the mass estimates of the common clusters overlapping 
in the P11 and P15 samples shown in Figure~\ref{norma}(c) gives
\begin{equation}
\log M_{\rm 500,P15}=(0.90\pm0.02)\log M_{\rm 500,P11}+(0.15\pm0.01).
\end{equation}
Combining the above two conversion equations, we get
\begin{equation}
\log M_{\rm 500,V09}=(0.95\pm0.03)\log M_{\rm 500,P11}+(0.18\pm0.02).
\label{v09p11}
\end{equation}

The H13 sample of the SZ clusters has even fewer overlaps with the
samples of V09 and P15. However, there are some common
clusters overlapping in the H13 and P11 samples. We compare their 
mass estimates as shown in Figure~\ref{norma}(d), and get
\begin{equation}
\log M_{\rm 500,P11}=(0.98\pm0.30)\log M_{\rm 500,H13}-(0.19\pm0.30).
\end{equation}
When it is combined with Equation~(\ref{v09p11}), we get the
conversion equation to the V09 standard as
\begin{equation}
\log M_{\rm 500,V09}=(0.93\pm0.29)\log M_{\rm 500,H13}+(0.00\pm0.29).
\end{equation}
The same holds for the clusters in the T14 sample. The best fit between
the T14 mass and the P11 mass as shown in Figure~\ref{norma}(e)
is
\begin{equation}
\log M_{\rm 500,P11}=(1.04\pm0.08)\log M_{\rm 500,T14}-(0.01\pm0.02),
\end{equation}
and the conversion to the V09 standard is through
\begin{equation}
\log M_{\rm 500,V09}=(0.99\pm0.08)\log M_{\rm 500,T14}+(0.17\pm0.03).
\end{equation}

\begin{figure}
\centering\resizebox{0.43\textwidth}{!}{\includegraphics{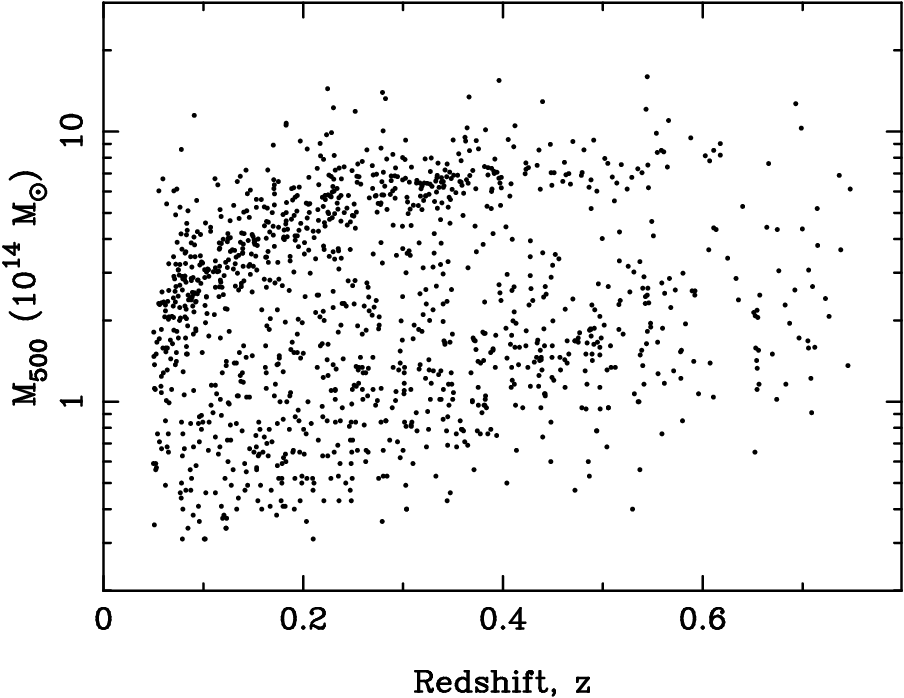}}
\caption{The masses for 1191 clusters in the composite calibration
  sample are in the range of 0.3 -- 20$\times10^{14}~M_{\odot}$, and the
  redshifts are distributed up to 0.75.
\label{mass_z}}
\end{figure}

The masses of clusters thus scaled in Figure~\ref{norma}(a)--(e) have 
excellent consistency, as shown in Figure~\ref{norma}(f), which shows a
scatter of only 0.1 dex, i.e. $\sigma_{\log M_{500}}=0.10$. Most of
the galaxy clusters from the V09, M10, P11, H13, and P15 samples have a
rescaled mass of $M_{500}>2\times10^{14}M_{\odot}$, while clusters in the
T14 sample often have a smaller mass of
$0.3\times10^{14}M_{\odot}<M_{500}<2\times10^{14}M_{\odot}$.

The rescaled masses of the galaxy clusters from the six samples will
be used to calibrate the optical mass proxy, which will be derived
from the SDSS data. Therefore we limit the composite sample to galaxy
clusters in the SDSS sky region, which must have a spectroscopic
redshift either in the NASA/IPAC Extragalactic Database or from the
SDSS spectroscopic observations \citep{dr12+15}. We also limit the
sample to a mass of $M_{500}>0.3\times10^{14}~M_{\odot}$ and in the
redshift range of $0.05<z<0.75$. The lower redshift limit is set due
to a photometry problem for very bright galaxies in the SDSS data, and
the redshift upper limit is considered for the reliable detection of
cluster galaxies in the SDSS data. We further visually inspect the
color
images\footnote{http://skyserver.sdss.org/dr12/en/tools/chart/navi.aspx}
and exclude nine clusters that are seriously contaminated by very
bright exposure-saturated stars. Finally we get 29, 64, 584, 50, 412,
and 370 clusters from the V09, M10, P11, H13, T14 and P15 samples,
respectively. After the overlapping common clusters are merged with
their masses estimated from uncertainty-weighted average, the
composite calibration sample consists of 1191 clusters, as listed in
Table~\ref{tab1}. The cluster radius $r_{500}$ are re-estimated from
the rescaled mass $M_{500}$ by using Equation~(\ref{m500def}). The
distributions of their masses and redshifts are shown in
Figure~\ref{mass_z}.

\subsection{Optical luminosity and richness of galaxy clusters}
\label{optlum}

We discriminate the cluster member galaxies and calculate the total
optical luminosity of clusters by using the SDSS spectroscopic and
photometric data following our previous procedures in
\citet[][hereafter WHL12]{whl12} but with a few improvements. The
cluster richness is defined from the optical luminosity, which can act
as the optical mass proxy for galaxy clusters.

\subsubsection{Discrimination of cluster member galaxies}

We only consider the bright member galaxies of $M^e_r(z)\le-20.5$.
Here, $M^e_r$ is the evolution-corrected absolute magnitude in the $r$
band, $M^e_r(z)=M_r(z)+Qz$. Assuming the passive evolution of a galaxy
population formed at $z_f=2$ \citep{lmg+06,cbh09} and adopting a
stellar population synthesis model \citep{bc03} with the initial mass
function of \citet{cha03} and solar metallicity, we obtain the value
of $Q\sim1.16$ at redshift $z<0.8$. The bright member galaxies of a
cluster can be discriminated around a brightest cluster galaxy (BCG) in
a small redshift range, i.e. a redshift slice.

To discriminate the cluster galaxies, the spectroscopic redshifts are
taken if they are available in the SDSS DR12 \citep{dr12+15};
otherwise, photometric redshifts (photo-$z$s hereafter) will be
used. For the objects classified as galaxies with only photo-$z$s, we
remove those with deblending problems and those marked as `SATURATED'
using the flags\footnote{(flags \& 0x20) = 0 and (flags \& 0x80000) =
  0 and ((flags \& 0x400000000000) = 0 or psfmagerr$_r <= 0.20$) and
  ((flags \& 0x40000) = 0)}, and also discard those objects with a
large photo-$z$ error $z{\rm Err}>0.08(1+z)$. These selection
procedures remove most of the contamination from stars and galaxies
with a bad photometry.

\begin{figure}
\centering\resizebox{0.4\textwidth}{!}{\includegraphics{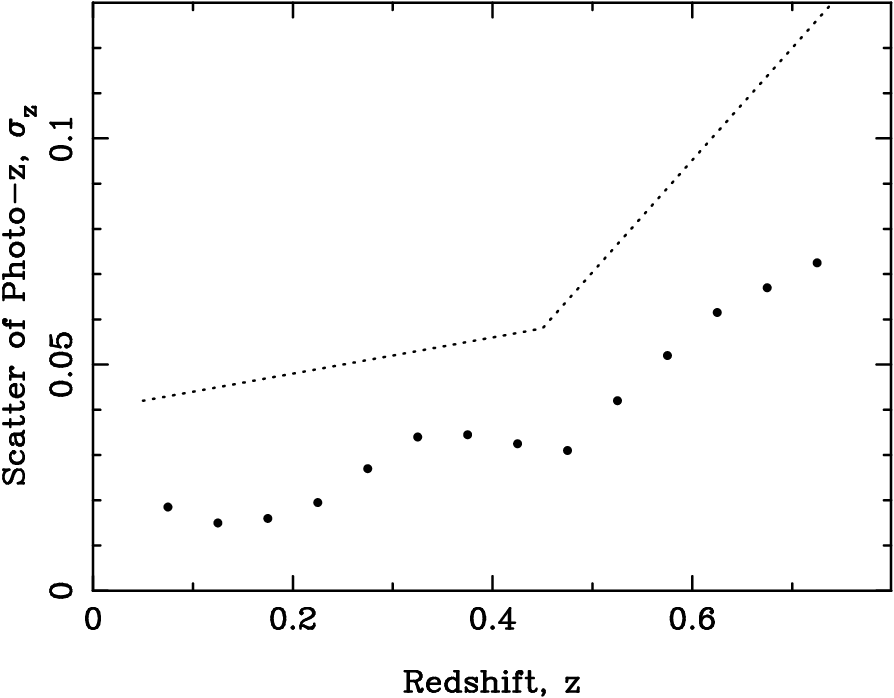}}
\caption{Data scatter $\sigma_z$ of photo-$z$ varies with redshift,
  obtained from the comparison of spectroscopic redshifts and
  estimated photo-$z$s of a large SDSS galaxies. The dotted line
  represents the half thickness $\Delta z$ of the photo-$z$ slice
  defined by Equation~(\ref{pzslice}) for discrimination of member
  galaxies.}
\label{photoz}
\end{figure}

For galaxies with spectroscopic redshifts, only those within a
redshift slice of $c\Delta z/(1+z)=2500$ km~s$^{-1}$ are considered as
member galaxies. Here $c$ is the speed of light, and $z$ is the
redshift of a cluster.

For most member galaxies, their photo-$z$s were estimated by optical
magnitudes in five bands of the SDSS and have an uncertainty much
larger than those of spectroscopic redshift \citep{cdt+07}. The data
scatter of photo-$z$ varies with redshift, as shown in
Figure~\ref{photoz}, and is smaller than 0.035 at a redshift of
$z<0.5$ but increases to $\sigma_z \sim 0.075$ at a redshift of
$z>0.7$.
Only galaxies within the redshift slice of $z\pm\Delta z$ are
considered as member galaxies of a cluster at a redshift of $z$. In
\citet{whl09} and \citet{whl12} the half thickness of the slice is
taken as $\Delta z=0.04\,(1+z)$, and the majority of the member
galaxies can be well discriminated for clusters with a redshift of
$z<0.42$. However, due to the large uncertainty of photo-$z$, such a
slice leads to incomplete recognizance of member galaxies for clusters
at higher redshifts. Here we adopt the photo-$z$ slice in the form of
\begin{equation}
  \Delta z=\left\{
\begin{array}{ll}
   0.04\,(1+z)           &     \mbox{for $z\leq0.45$}\\
   0.248\,z-0.0536       &     \mbox{for $z>0.45$}
\end{array}
\right. \,,
\label{pzslice}
\end{equation}
which is about 1.75--2.00\,$\sigma_z$ and by which most cluster member
galaxies can be identified up to a high redshift. Such a larger
photo-$z$ slice leads to a slightly higher completeness for
identification of member galaxies but introduces more contamination.

Compared to \citet{whl12}, we improve the discrimination of member
galaxies in three aspects: (a) spectroscopic redshifts are considered
for member galaxies; (b) the evolution of absolute magnitudes is
improved from $Q = 1.62\pm0.30$ \citep{bhb+03} to the current value of
$Q=1.16$; and (c) the thickness of the photo-$z$ slice at high redshifts
is enlarged according to the photo-$z$ scatter.

\subsubsection{Total $r$-band luminosity of cluster member galaxies
  and background estimation}

In general for each known cluster, the total $r$-band luminosity can
be calculated by summing the luminosity of recognized member
galaxies. The total luminosity is measured in units of $L^{\ast}$
which is the evolved characteristic luminosity of galaxies in the
$r$-band defined by $L^{\ast}(z)=L^{\ast}(0)10^{0.4Qz}$, here again
$Q=1.16$.

However, the contamination by background field galaxies has to be
subtracted. To estimate the local background, we follow the method
analogous to \citet{pbb+04}. For each cluster, we take the BCG as the
center of a galaxy cluster, and then divide the annulus between 2--4
Mpc from the BCG into 48 sectors with equal areas. Within the same
magnitude limit and the same photo-$z$ slice, the field galaxies in
these sectors fainter than the second brightest cluster member
galaxies are used to calculate the background. We get the total
$r$-band luminosity of bright galaxies in each sector, and estimate
the mean and deviation $\sigma$. The sectors with a luminosity
obviously deviating from the mean by more than $3\sigma$ are
discarded, and the mean is recalculated for the local background which
is scaled and subtracted from the total luminosity.

\begin{figure*}
\centering
  \epsscale{0.7}
\includegraphics[width=0.3\textwidth]{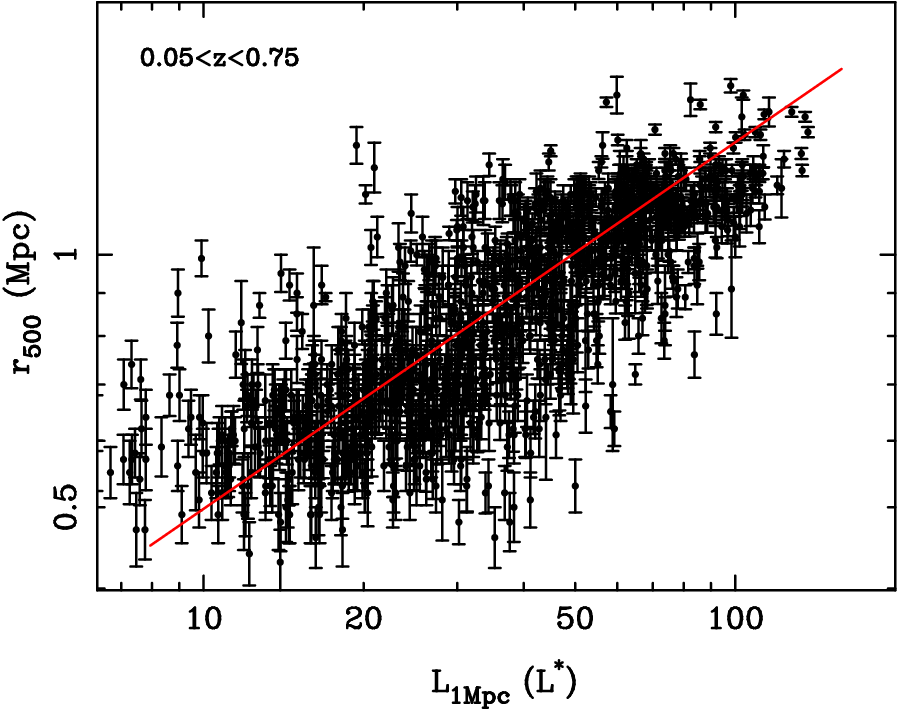} \hspace{3mm}
\includegraphics[width=0.3\textwidth]{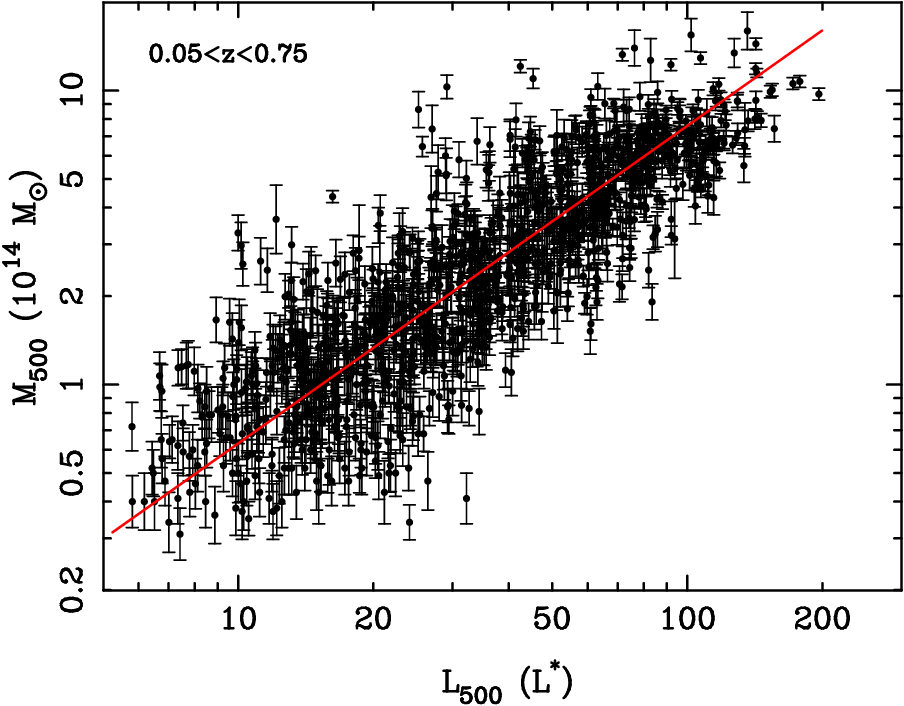} \hspace{3mm}
\includegraphics[width=0.3\textwidth]{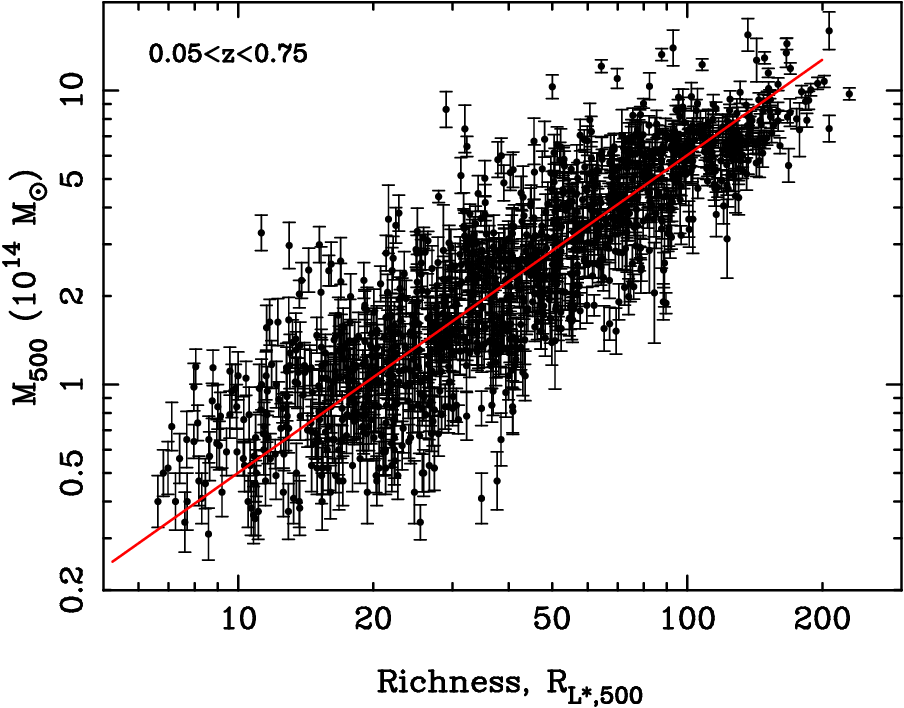}\\[1mm]
\includegraphics[width=0.3\textwidth]{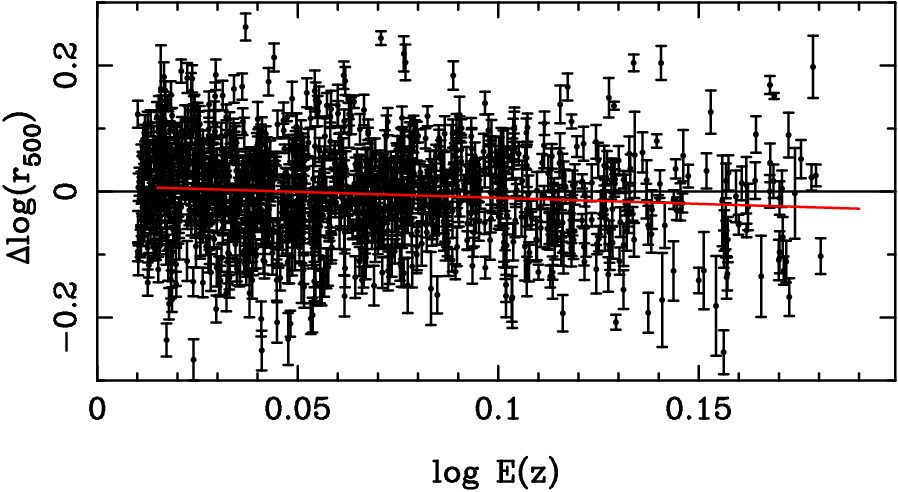} \hspace{3mm}
\includegraphics[width=0.3\textwidth]{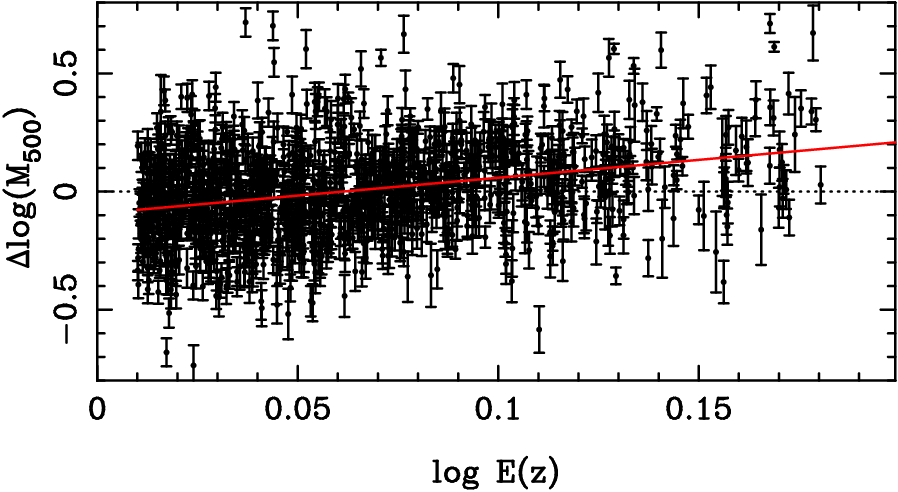} \hspace{3mm}
\includegraphics[width=0.3\textwidth]{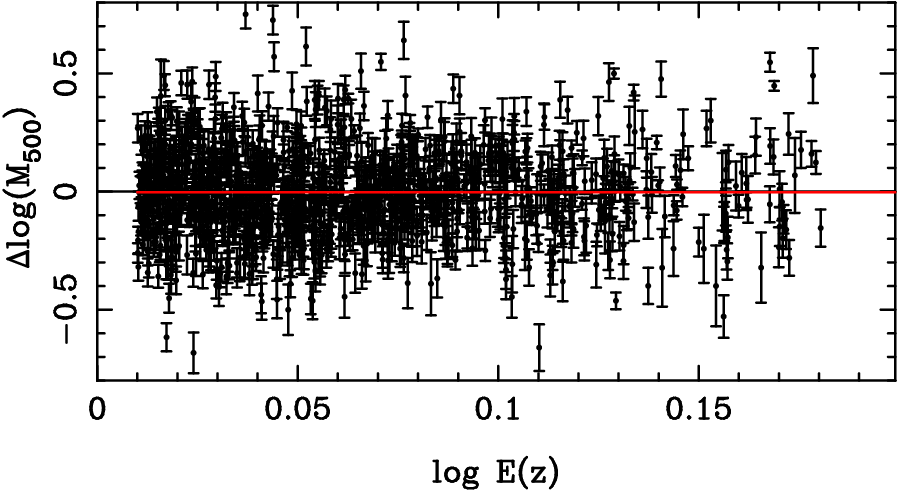}
\caption{Upper panels are the scaling relations between cluster
  parameters derived from the optical data (the X-axis) and cluster
  radius or mass (the Y-axis) for the 1191 clusters from the composite
  calibration sample. The data scatters are plotted in the lower
  panels against $\log E(z)$ to check their possible evolution.}
\label{scaleXO}
\end{figure*}

\subsubsection{Scaling relations between the optical data and mass}
\label{calib}

The goal here is to establish the scaling relation between the cluster
mass and the luminosity or richness derived from optical data. As
shown by \citet{whl12}, if there are merely optical survey data for a
cluster, a few steps are needed to derive the richness from optical
data: (1) discriminate cluster member galaxies and get their
redshifts; (2) get the optical luminosity $L_{\rm 1Mpc}$ from member
galaxies within 1~Mpc, and then derive the cluster radius $r_{500}$;
and (3) get the optical luminosity $L_{500}$ within $r_{500}$, and
then the richness defined as $R_{L\ast,500} = L_{500}/L^\ast$.

First, to estimate $r_{500}$ from optical data, we have to establish
the relation between $r_{500}$ and an optical observable that is the
total $r$-band luminosity $L_{\rm 1Mpc}$ (in units of $L^{\ast}$, with
background subtracted) of member galaxies within 1~Mpc from the
cluster center. As shown in the left panels of Figure~\ref{scaleXO},
we find that the values of $r_{500}$ (in Mpc) in Table~\ref{tab1} are
well correlated with $L_{\rm 1Mpc}$. The uncertainty of $L_{\rm 1Mpc}$
comes from the incompleteness of member galaxies and residual
contamination of background galaxies, which is about 10\% for rich
clusters \citep{whl09}. By taking a typical uncertainty of 15\% on
$L_{\rm 1Mpc}$ for all clusters, we find the best fit to the data as
\begin{equation}
\log r_{500}=(0.45\pm0.01)\log L_{\rm 1Mpc}-(0.76\pm0.01).
\label{r500}
\end{equation}
The slope obtained here is consistent with that in \citet{whl12}. 

Some authors have shown that the scaling relation between $M_{500}$
(or $r_{500}$) and an optical observable (e.g., richness) may evolve
with redshift \citep{lms04,ac14,wls15}, for example, due to higher
concentration of the light-to-mass profiles at lower redshifts
\citep{ggi+13}. On the other hand, bright member galaxies of
$M_e\le-20.5$ are incomplete in the SDSS photometric data at $z>0.45$,
so that the total luminosity could be underestimated.  The 1191
clusters in the composite calibration sample are in a very large
redshift range, allowing us to clarify any redshift evolution of the
scaling relations. For the scaling relation $r_{500}$--$L_{\rm
  1Mpc}$, we calculate the deviation of $r_{500}$ from the relation:
$\Delta\log r_{500}=\log r_{500}-(0.45\,\log L_{\rm 1Mpc}-0.76)$ and
find that $\Delta\log r_{500}$ is only slightly evolving with the
redshift in the form of (see the lower left panel in
Figure~\ref{scaleXO})
\begin{equation}
\Delta\log r_{500}=(-0.20\pm0.06)\log E(z)+(0.01\pm0.01),
\end{equation}
where $E(z)=\sqrt{\Omega_{\Lambda}+\Omega_m(1+z)^3}$.
Including this small correction, the radius $r_{500}$ is related
to $L_{\rm 1Mpc}$ and $z$ by
\begin{eqnarray}
\log r_{500}&=&(0.45\pm0.01)\log L_{\rm 1Mpc}-(0.75\pm0.01)\nonumber\\
&-&(0.20\pm0.06)\log E(z).
\label{r500z}
\end{eqnarray}

We calculate the total $r$-band luminosity $L_{500}$ (also in units of
$L^{\ast}$, with background subtracted) of member galaxies within the
radius of $r_{500}$ for the 1191 clusters by using the SDSS data. As
shown in the upper middle panel of Figure~\ref{scaleXO}, there exists
excellent correlation between $L_{500}$ and the cluster mass $M_{500}$
(in units of $10^{14}~M_{\odot}$), and the best fit is
\begin{equation}
\log M_{500}=(1.08\pm0.02)\log L_{500}-(1.28\pm0.02).
\end{equation}
Again, to check for the possible evolution with redshift, the deviations
of $M_{500}$ from the above fitted relation, $\Delta\log M_{500}=\log
M_{500}-(1.08\,\log L_{500}-1.28)$, are plotted against $E(z)$ in the
middle lower panel of Figure~\ref{scaleXO}, which shows a correlation
in the form of
\begin{equation}
\Delta\log M_{500}=(1.51\pm0.14)\log E(z)-(0.09\pm0.01),
\end{equation}
Though the total $r$-band luminosity $L_{500}$ in principle may
slightly be underestimated for clusters of $z>0.45$ (i.e.,$\log
E(0.45)=0.104$), the $\Delta\log M_{500}$ has to be so corrected.
We then can define the richness of the galaxy cluster from the
total $r$-band luminosity $L_{500}$, which can act as a
well-calibrated mass proxy in the optical band, as 
\begin{equation}
R_{L\ast,500}=L_{500}\,E(z)^{1.40}.
\end{equation}
Such a richness is independent of redshift and is valid in the
redshift range of $0.05<z<0.75$, as shown in the right panels of
Figure~\ref{scaleXO}. The thus-calculated richness $R_{L\ast,500}$ for
galaxy clusters in the composite calibration sample is listed in
Table~\ref{tab1}. The scaling relation between the cluster mass and
the richness is
\begin{equation}
\log M_{500}=(1.08\pm0.02)\log R_{L\ast,500}-(1.37\pm0.02),
\label{richm500}
\end{equation}
which is consistent with but more accurate than the mass--richness
relation obtained in our previous work \citep{whl12} and the
$M_{500}$--$\lambda$ relation given by \citet{rkr+12}. The scatter of
mass values is about 0.17 dex, i.e. $\sigma_{\log M_{500}}=0.17$.
Taking out the inherent scatter $\sigma_{\log M_{500}}=0.10$ of the
rescaled masses of the composite calibration sample, the mass
uncertainty estimated by $R_{L\ast,500}$ should be $\sigma_{\log
  M_{500}}=0.14$.

Because clusters in the composite calibration sample are distributed
in a wide range of mass and redshift, the scaling relations we
derived above should be valid for clusters in the SDSS in the redshift
ranges of $0.05<z<0.75$ with a mass of
$M_{500}>0.3\times10^{14}~M_{\odot}$.

\begin{table*}
\begin{center}
  \caption{Comparison of data scatter for masses estimated from different
    richnesses in different cluster catalogs.}
\label{tab4}
\begin{tabular}{rrccc}
\hline
\mc{1}{c}{Cluster catalog} &\mc{1}{c}{Redshift range} &\mc{1}{c}{Number of} &\mc{1}{c}{$\sigma_{\log M_{500}}$} & 
\mc{1}{c}{$\sigma_{\log M_{500}}$} \\
\mc{1}{c}{} &\mc{1}{c}{for matching} &\mc{1}{c}{matched clusters} &\mc{1}{c}{scaled from ref.} & \mc{1}{c}{from this work} \\
\hline
maxBCG       &$0.10<z<0.30$ &325 & 0.29 & 0.20 \\
GMBCG        &$0.10<z<0.37$ &502 & 0.27 & 0.19 \\
AMF          &$0.05<z<0.38$ &402 & 0.26 & 0.20 \\
WHL12        &$0.05<z<0.42$ &799 & 0.24 & 0.20 \\
redMapper    &$0.08<z<0.55$ &610 & 0.17 & 0.18 \\
CAMIRA       &$0.1<z<0.6$   &647 & 0.18 & 0.19 \\
Yang et al.  &$z<0.2$       &195 & 0.56 & 0.19 \\
Tempel et al.&$z<0.2$       &149 & 0.23 & 0.19 \\
This paper   &$0.05<z<0.75$ &1191& --   & 0.20 \\
\hline
\end{tabular}
\end{center}
\end{table*}

\begin{figure}
\epsscale{1.}
\plotone{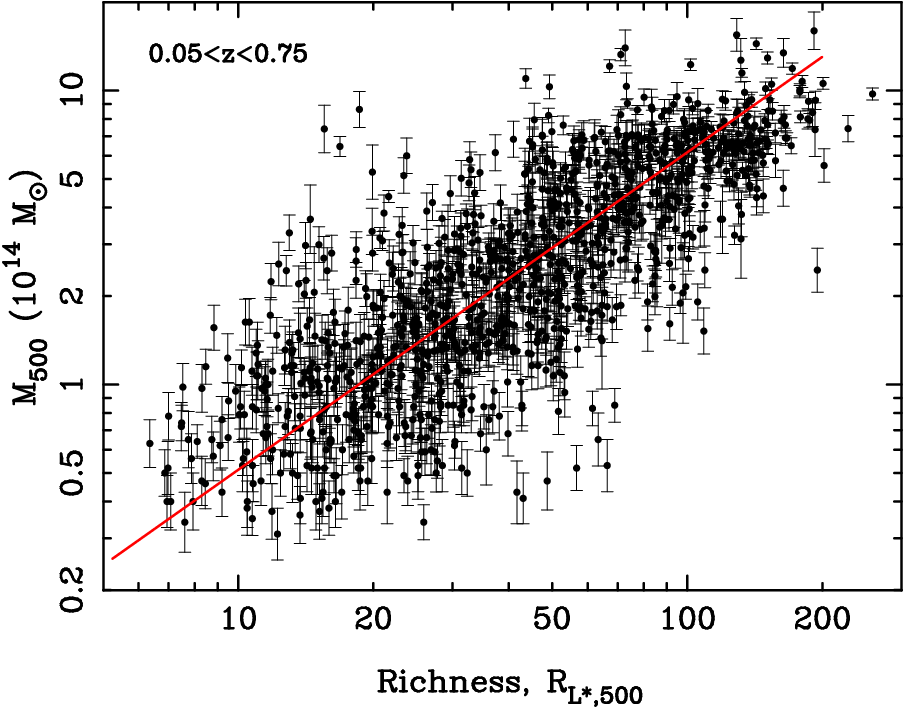}
\vspace{3mm}
\plotone{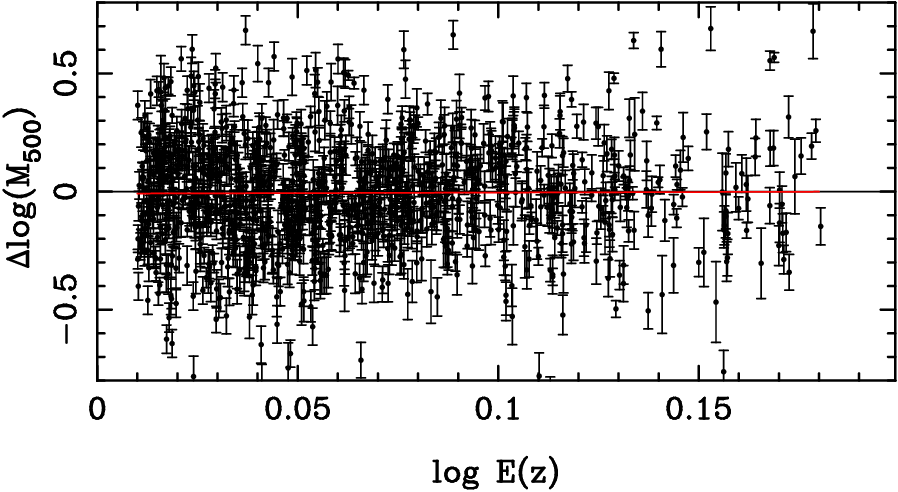}
\caption{Correlation between $M_{500}$ and the directly optically
  estimated $R_{L\ast,500}$ for the 1191 clusters in
  Table~\ref{tab1}. Different from the right panel of
  Figure~\ref{scaleXO}, the values of $r_{500}$ and $R_{L\ast,500}$
  are both estimated from merely optical data.
\label{scale_MR}}
\end{figure}

\subsection{Goodness of optical mass proxies}

How good is the richness $R_{L\ast,500}$ estimated from just the optical
data (i.e., $r_{500}$ is unknown previously) acting as a mass proxy?
To get a quantitative estimate, we estimated the $L_{\rm 1Mpc}$ of 1191
clusters in the composite calibration sample, and then get $r_{500}$
from the scaling relation in Equation~(\ref{r500z}) so that $L_{500}$
and hence $R_{L\ast,500}$ can be estimated from the member galaxies
within $r_{500}$. Obviously the uncertainty of estimated $r_{500}$ is
propagated to $L_{500}$ and then $R_{L\ast,500}$. The plot for
$M_{500}$ and $R_{L\ast,500}$ in Figure~\ref{scale_MR} shows a mass 
scatter in $0.20$ dex, i.e. $\sigma_{\log M_{500}}=0.20$. Taking out the
inherent scatter of $\sigma_{\log M_{500}}=0.10$ for the masses
estimated from the X-ray and SZ observations, the mass scaled by
$R_{L\ast,500}$ should have a scatter of $\sigma_{\log M_{500}}=0.17$.

Can the richness in other cluster catalogs be used as a mass
proxy? If so, how good is it? To find out the answer to these
questions, we investigate the optical richness based on photometric
data in the cluster catalogs of the maxBCG \citep{kma+07}, the
Gaussian Mixture Brightest Cluster Galaxy \citep[GMBCG,][]{hmk+10},
the adaptive matched filter \citep[AMF,][]{spd+11}, the WHL12
\citep{whl12}, the red-sequence Matched-filter Probabilistic
Percolation \citep[redMapper,][]{rrb+14}, the Cluster finding
Algorithm based on Multi-band Identification of Red-sequence galaxies
\citep[CAMIRA,][]{ogu14}, and the richness based on spectroscopic data
by \citet{ymv+07} and \citet{ttg+14}.

The richness is defined as the total number of galaxies in the
catalogs of maxBCG, GMBCG, and redMapper, or defined as the total
luminosity of member galaxies in the catalogs of the AMF and WHL12.
\citet{ymv+07} adopted the characteristic luminosities or the
characteristic stellar masses as the halo mass proxy of galaxy
group/clusters, and here we take the halo mass for comparison.
\citet{ttg+14} provided the dynamical mass estimates based on
radial velocity dispersion of member galaxies, which we will
take for comparison.

We cross-match the galaxy clusters in Table~1 with clusters in the
above cluster catalogs, and get the matched clusters within a
projected distance of $r_{500}$ and a redshift difference of 0.05. If
more than one cluster in a catalog matches with a cluster in Table~1,
we take the one with the minimum projected distance. Note that for
different cluster catalogs, we take the redshift ranges up to their
redshift distribution peaks. As shown in Table~\ref{tab4}, we get
different numbers of matched clusters for different catalogs. We
estimate the cluster mass by using each richness and get the scatter
of the mass. For comparison, the mass scatter for the same sample of
matched clusters for each catalog is also calculated by using our
newly defined richness. We find that the richness defined in this
paper and those in redMapper and CAMIRA are equally good to act as
optical mass proxy. The improvement in this paper is that the newly
defined richness has the redshift evolution corrected and is good for
a wide range of redshifts up to $z\sim0.75$.

\section{Update of the WHL12 cluster catalog}

We identified 132,684 galaxy clusters by using the photo-$z$ of
galaxies from SDSS DR8 \citep{whl12}. Clusters were identified if they
have a richness $\ge 12$ and the number of member galaxies $N_{200}
\ge 8$ within a radius of $r_{200}$. Here, $r_{200}$ is the radius
within which the mean density of a cluster is 200 times of the
critical density of the universe. The cluster richness $R_{L\ast}$ was
defined to be the $r$-band total luminosity (in units of $L^{\ast}$)
of member galaxies with $M^e_r\le-20.5$ within $r_{200}$.

Two reasons lead us to update the WHL12 cluster catalog. First, the
latest SDSS DR12 \citep{dr12+15} released spectroscopic redshifts for
about 2.30 million galaxies, of which about 1.35 million are luminous
red galaxies (LRGs). In general LRGs are massive galaxies, and many of
them are the BCGs or the bright member galaxies of clusters in the
WHL12. We can update these spectroscopic redshifts for most of the
WHL12 clusters. Second, the newly defined richness can act as an
excellent mass proxy, and should be calculated for all clusters in the
WHL12 catalog. In addition, by using the SDSS DR12 data, we can find
rich galaxy clusters at high redshifts.

\subsection{Update on spectroscopic redshifts and the BCGs}

In general, we take the spectroscopic redshift of the BCG $z_{\rm
  s,BCG}$ as the redshift of a cluster. SDSS DR12 has spectroscopic
redshifts for about 2.30 million galaxies \citep{dr12+15}. Simple
cross-matching the BCGs of the WHL12 clusters with the galaxies with
spectroscopic data in the SDSS DR12 gives the redshifts for 84,760
(64\%) clusters.

We then try to get spectroscopic redshifts of bright member
galaxies. For clusters with available $z_{\rm s,BCG}$, we consider
only galaxies within $r_{500}=2/3\,r_{200}$ from the BCG and within
2500~km~s$^{-1}$ from $z_{\rm s,BCG}$ in the rest frame. Here
$r_{200}$ can be found in Table~1 of \citet{whl12}. For clusters
without $z_{\rm s,BCG}$, we consider the galaxies within
$r_{500}=2/3\,r_{200}$ from the BCGs and within a redshift difference
of $0.03(1+z)$ from $z_{\rm p}$, the photo-$z$ of clusters in
\citet{whl12}. This small redshift difference is about 2.0--2.5 times
the uncertainty of the cluster photo-$z$. The mean spectroscopic
redshifts of the member galaxies (including the BCG) is regarded to be
the spectroscopic redshift of a cluster. Through this process, we get
spectroscopic redshifts for 95,684 clusters, which is about 72\% in
the WHL catalog. In the SDSS spectroscopic survey region, 85\% of the
WHL clusters have spectroscopic redshifts obtained. Clusters located
at low Galactic latitudes have no spectroscopic redshifts yet.

The position and the BCG have been updated for some clusters as well.
In \citet{whl12}, all objects with the deblending problems or marked
as ``SATURATED'' (see the flags in Section~\ref{optlum}) were removed
from the photometric data in the procedure for identifying galaxy
clusters. The true BCGs that have such a flag or catastrophic
photo-$z$ have now been included by using the spectroscopic
redshifts. If a galaxy is brighter than the BCG of an original WHL12
cluster within $r_{200}$ and has a redshift difference of $0.03(1+z)$
and is surrounded by other member galaxies in the SDSS color image, we
take it to be the new BCG of the cluster. We hence update the BCGs for
3128 clusters.

\begin{figure}
\epsscale{1.}
\plotone{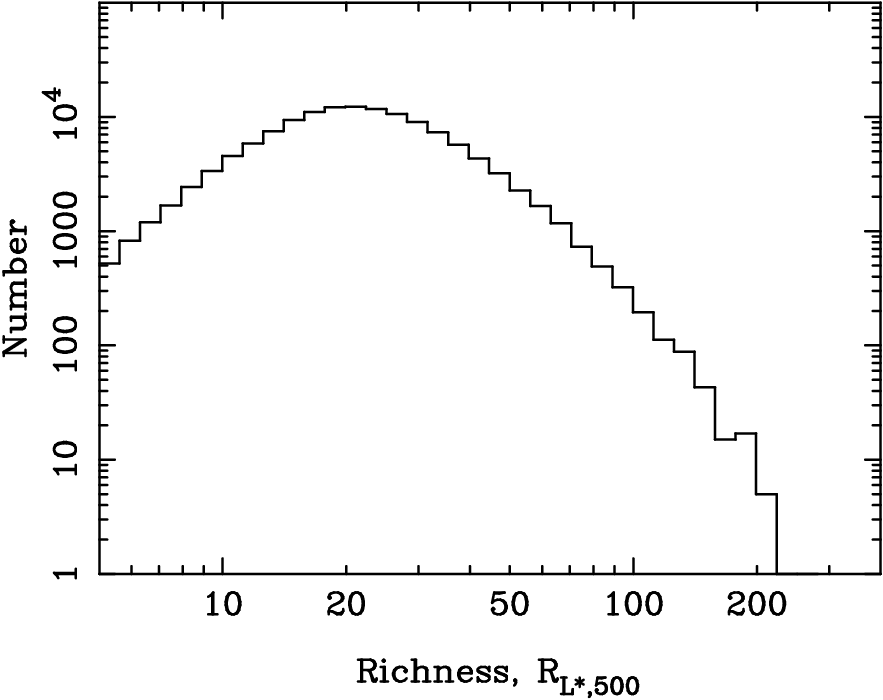}
\caption{Distribution of the newly defined richness $R_{L\ast,500}$ for
  the WHL12 clusters.}
\label{newwhl}
\end{figure}

\begin{table*}
{\footnotesize
\begin{center}
\caption{The WHL12 cluster catalog with updated parameters}
\label{tab2}
\begin{tabular}{ccccrcccccccc}
\hline
\mc{1}{l}{Cluster name} &\mc{1}{c}{R.A.}  &\mc{1}{c}{Decl.}  &\mc{1}{c}{$z_p$}  & \mc{1}{c}{$z_s$} & 
\mc{1}{c}{$r_{\rm BCG}$} & \mc{1}{c}{$r_{200}$} & \mc{1}{c}{$R_{L*}$} & \mc{1}{l}{$N_{200}$} & 
\mc{1}{c}{$r_{500}$} &\mc{1}{c}{$R_{L\ast,500}$} & \mc{1}{c}{$N_{500,\rm sp}$} & \mc{1}{l}{$N_{500}$}\\
        &\mc{1}{c}{(deg)} &\mc{1}{c}{(deg)} &     &     &    &  (Mpc)  &    &   & (Mpc) &  & & \\
(1) & (2) & (3) & (4) & (5) & (6) & (7) & (8) & (9) & (10) & (11) & (12) & (13)\\
\hline  
WHL J000000.6$+$321233 & 0.00236 &  32.20925& 0.1274 &$-1.0000$& 14.92 & 1.72 & 70.63 & 24 & 1.01 & 68.28 &  0 & 19\\
WHL J000002.3$+$051718 & 0.00957 &   5.28827& 0.1696 &  0.1694 & 16.20 & 0.94 & 17.48 &  9 & 0.72 & 24.46 &  1 &  9\\
WHL J000003.1$-$033245*& 0.01276 &$-3.54578$& 0.5968 &  0.5973 & 20.14 & 1.00 & 19.19 & 11 & 0.83 & 49.33 &  2 & 13\\
WHL J000003.3$+$311354 & 0.01377 &  31.23175& 0.5428 &  0.5293 & 20.17 & 0.87 & 14.27 &  8 & 0.63 & 25.43 &  2 &  9\\
WHL J000003.5$+$314708 & 0.01475 &  31.78564& 0.0932 &  0.0916 & 15.18 & 0.94 & 16.97 &  9 & 0.69 & 21.14 &  1 & 11\\
WHL J000004.7$+$022826 & 0.01945 &   2.47386& 0.4179 &$-1.0000$& 19.32 & 0.95 & 13.71 & 10 & 0.54 &  6.60 &  0 &  3\\
WHL J000005.5$+$354610 & 0.02303 &  35.76957& 0.4762 &$-1.0000$& 19.59 & 0.91 & 15.58 &  9 & 0.63 & 19.13 &  0 &  6\\
WHL J000006.0$+$152548 & 0.02482 &  15.42990& 0.1656 &  0.1731 & 16.60 & 1.13 & 23.53 & 19 & 0.79 & 19.53 &  2 & 12\\
WHL J000006.3$+$221220 & 0.02643 &  22.20558& 0.3985 &$-1.0000$& 19.36 & 0.84 & 12.73 & 11 & 0.52 & 11.92 &  0 &  7\\
WHL J000006.6$+$100648 & 0.02755 &  10.11333& 0.3676 &  0.3747 & 19.07 & 0.93 & 16.73 & 13 & 0.70 & 23.52 &  2 & 12\\
\hline
\end{tabular}
\end{center}
{Note.
Column (1): Cluster name, the name with `*' means the BCG is updated; 
Column (2) and (3): R.A. (J2000) and Decl. (J2000) of cluster; 
Column (4): photometric redshift of cluster; 
Column (5): spectroscopic redshift of cluster, $-1.0000$ means not available; 
Column (6): $r$-band magnitude of the BCG;
Column (7): original $r_{200}$ of cluster (Mpc);
Column (8): original richness;
Column (9): original number of member galaxy within $r_{200}$;
Column (10): cluster radius $r_{500}$ derived from optical luminosity;
Column (11): newly defined cluster richness, $R_{L\ast,500}$;
Column (12): number of member galaxies with spectroscopic redshifts within $r_{500}$;
Column (13): number of all member galaxies within $r_{500}$.
Column (4) and (7)--(9) are from \citet{whl12}, Column (10)--(13) are derived in this paper.\\
(This table is available in its entirety in a machine-readable form in the online journal. 
A portion is shown here for guidance regarding the form and content.)
}
}
\end{table*}

\begin{table*}
{\footnotesize
\begin{center}
\caption{Newly identified rich clusters at high redshifts}
\label{tab3}
\begin{tabular}{ccccccccr}
\hline
\mc{1}{l}{Cluster name} &\mc{1}{c}{R.A.}  &\mc{1}{c}{Decl.}  &\mc{1}{c}{$z$}  & \mc{1}{c}{$r_{\rm BCG}$} 
& \mc{1}{c}{$r_{500}$} & \mc{1}{c}{$R_{L\ast,500}$} & \mc{1}{l}{$N_{500,\rm sp}$} & \mc{1}{l}{$N_{500}$}\\
              &\mc{1}{c}{(deg)} &\mc{1}{c}{(deg)} &                &                    
&     (Mpc)            &                       &                   &  \\
(1) & (2) & (3) & (4) & (5) & (6) & (7) & (8) & (9)\\
\hline  
WH J000004.2$+$021941 & 0.01743 &   2.32800 & 0.6443 & 20.15 & 0.69 & 27.88 &  1 &  6\\
WH J000004.3$-$091120 & 0.01809 & $-9.18902$& 0.6028 & 19.95 & 0.70 & 34.93 &  1 &  7\\
WH J000005.9$+$100144 & 0.02452 &  10.02896 & 0.6712 & 20.46 & 0.65 & 27.84 &  1 &  6\\
WH J000006.8$+$195244 & 0.02824 &  19.87897 & 0.4771 & 19.88 & 0.62 & 22.75 &  2 &  7\\
WH J000007.7$+$234443 & 0.03208 &  23.74541 & 0.5285 & 19.84 & 0.65 & 27.25 &  2 &  7\\
WH J000008.2$+$263228 & 0.03423 &  26.54098 & 0.5462 & 20.02 & 0.64 & 26.35 &  1 &  8\\
WH J000008.6$+$053257 & 0.03568 &   5.54922 & 0.5742 & 20.05 & 0.73 & 34.66 &  3 & 10\\
WH J000009.9$-$072520 & 0.04132 & $-7.42227$& 0.5847 & 20.78 & 0.62 & 24.19 &  1 &  8\\
WH J000010.4$+$041038 & 0.04318 &   4.17714 & 0.6954 & 21.29 & 0.58 & 24.96 &  1 &  6\\
WH J000013.5$-$022138 & 0.05613 & $-2.36069$& 0.4959 & 19.94 & 0.61 & 22.55 &  2 &  7\\
\hline
\end{tabular}
\end{center}
{Note.
Column (1): cluster name;
Column (2) and (3): R.A. (J2000) and Decl. (J2000) of cluster;
Column (4): spectroscopic redshift of cluster;
Column (5): $r$-band magnitude of the BCG;
Column (6): cluster radius $r_{500}$ derived from optical luminosity $L_{\rm 1~Mpc}$;
Column (7): cluster richness, $R_{L\ast,500}$;
Column (8): number of member galaxies with spectroscopic redshifts within $r_{500}$;
Column (9): number of all member galaxies within $r_{500}$.\\
(This table is available in its entirety in a machine-readable form in the online journal. 
A portion is shown here for guidance regarding the form and content.)
}
}
\end{table*}

\subsection{Update on cluster richness}

By using the scaling relations we obtained in Section~\ref{calib}, we
update the richness of galaxy clusters in the WHL12 catalog via
the following steps:
\begin{center}
$L_{\rm 1Mpc}\rightarrow r_{500} \rightarrow L_{500}\rightarrow R_{L\ast,500}\;\;.$
\end{center}
That is to say, we have to estimate $r_{500}$ from $L_{\rm 1Mpc}$, and
then get $L_{500}$ and $R_{L\ast,500}$. 

The updated parameters are listed in Table~\ref{tab2}, where the
cluster names, coordinates of the BCG, cluster redshift, and the
$r$-band magnitude, as well as the newly estimated radius $r_{500}$
and newly defined richness $R_{L\ast,500}$ are listed. The number of
member galaxies $N_{500}$ within $r_{500}$ and the number of these
galaxies with spectroscopic redshifts $N_{500,\rm sp}$ are given for
reference. The distribution of $R_{L\ast,500}$ is shown in
Figure~\ref{newwhl}; it has a peak around $R_{L\ast,500}=20$.  We
noticed that 5183 clusters in the WHL12 catalog has a low new richness
of $R_{L\ast,500}<8$.

\begin{figure}
\epsscale{1.}
\plotone{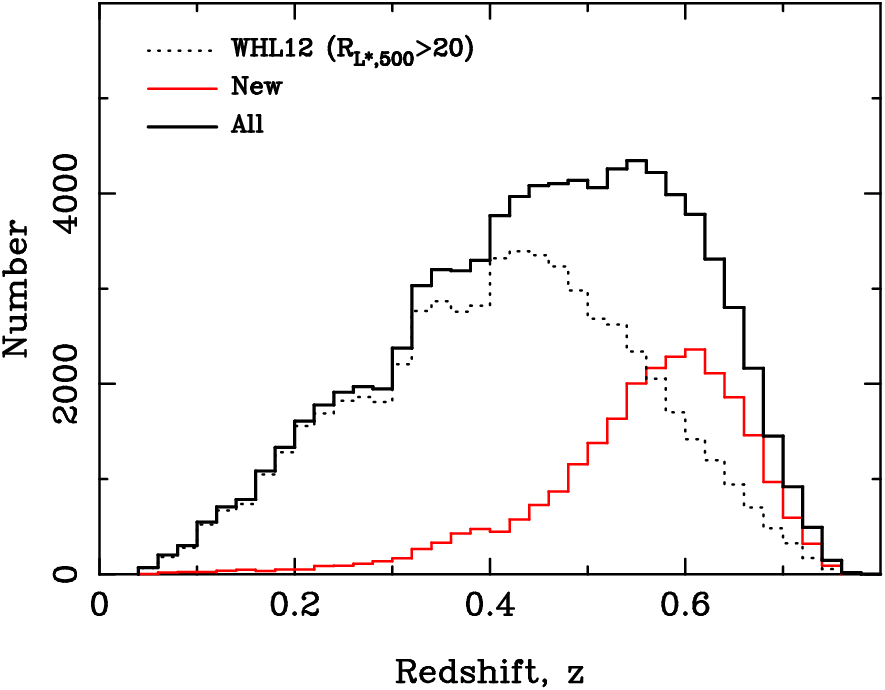}
\caption{Redshift distribution of the newly identified clusters, compared with
the WHL12 clusters of $R_{L\ast,500}\geq20$ in the SDSS spectroscopic survey region.
\label{histzall}}
\end{figure}

\begin{figure}
\centering
\epsscale{0.8}
\includegraphics[width=0.45\textwidth,height=0.22\textwidth]{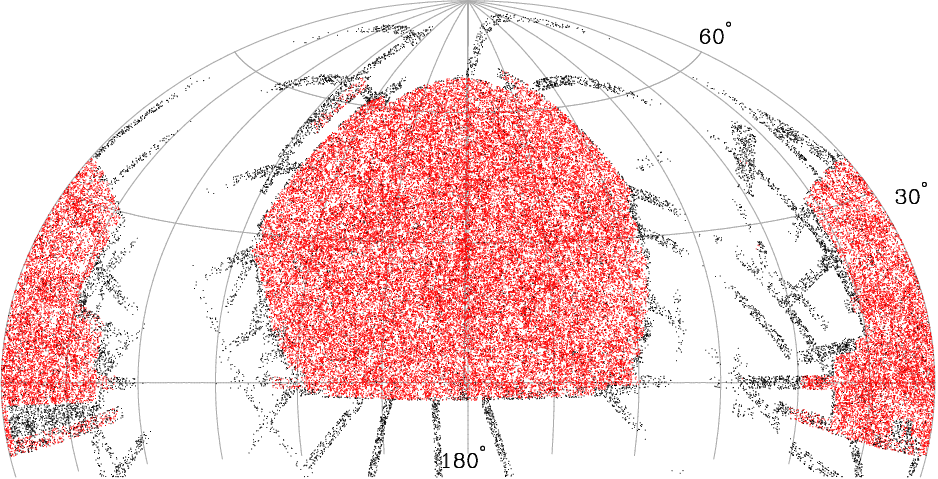} 
\includegraphics[width=0.45\textwidth,height=0.25\textwidth]{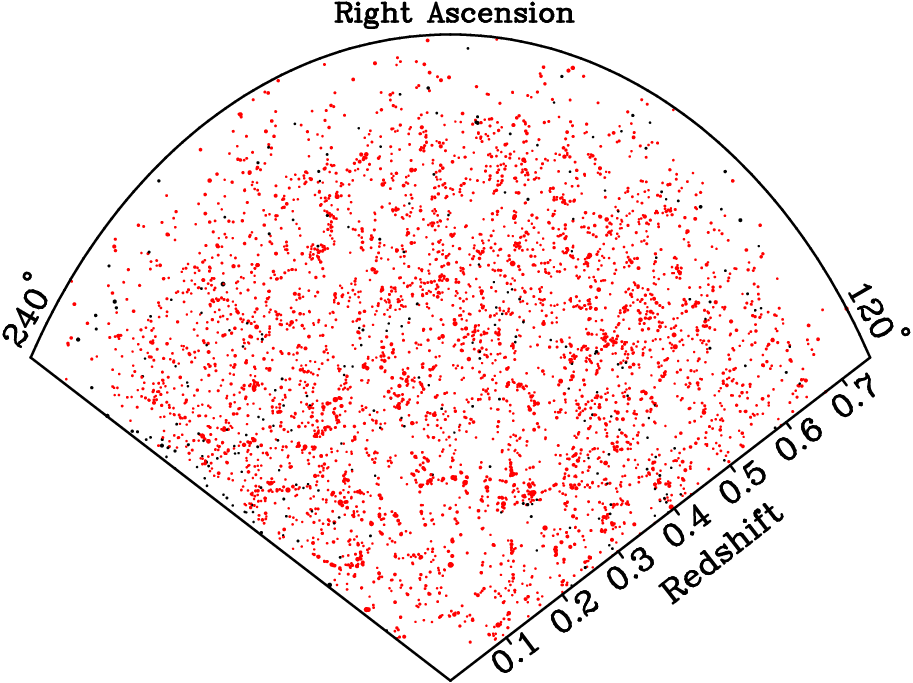}
\caption{Sky coverage of rich clusters of $R_{L\ast,500}\ge20$ (upper
  panel) and their radial distribution in a slice $0^{\circ}\leq {\rm
    Decl.}  \leq 5^{\circ}$ (lower panel). The red points are clusters
  with spectroscopic redshifts. The dot size in the lower panel is
  scaled by $\sqrt{R_{L\ast,500}}$.}
\label{sky}
\end{figure}

\subsection{Complementary galaxy clusters at high redshifts}
\label{higzcl}

Identification of galaxy clusters at high redshifts was incomplete due
to the large uncertainty of the photometric redshift, so that the the
number of clusters in the WHL12 catalog drops down at redshifts of
$z>0.42$ \citep[][see also Figure~\ref{histzall}]{whl12}. Here we
identify complementary rich clusters around bright galaxies which have
spectroscopic redshifts measured in the SDSS DR12, with following
procedures:

1. Get the spectroscopic redshift of a cluster candidate. Each bright
galaxy of $M_e\le-20.5$ with a spectroscopic redshift is assumed to be
a member galaxy of a cluster candidate. All galaxies within $0.5$ Mpc
from a given galaxy and with a velocity difference in the rest frame
less than 2500~km~s$^{-1}$ are taken to be the member galaxies of the
cluster candidate. Then the redshift of the cluster is defined as the
median value of the spectroscopic redshifts of these galaxies.

2. Determine the BCG. Generally, the true BCG, because it is the
brightest, should have been observed in the spectroscopic survey
already. However, the BCGs in some clusters may be not observed
spectroscopically because of their blue colors \citep{cae+99,psp+11}
or the fiber collision limitation in the SDSS spectroscopic
survey. The BCG candidates are the galaxies within 0.5~Mpc and with a
velocity difference less than 2500~km~s$^{-1}$ from the target galaxy
or with a photo-$z$ difference of $0.03(1+z)$.  The small slice of
photo-$z$ is used here for the BCG selection to reduce contamination
from field galaxies. The BCG of a cluster candidate is recognized as
the brightest galaxy of these galaxies.

3. Calculate the richness and judge a rich cluster. After we have the
redshift of a cluster candidate and also the recognized BCG, we then 
find the bright member galaxies in the redshift slice defined by
Equation~\ref{pzslice}, as described in Section~\ref{calib}, and then
calculate $L_{\rm 1Mpc}$ to estimate $r_{500}$, and then get $L_{500}$
and finally $R_{L\ast,500}$. We only identify rich clusters with a
richness of $R_{L\ast,500}\ge20$ (beyond the peak in Figure~\ref{newwhl})
and the number of bright galaxies within $r_{500}$, $N_{\rm 500}\ge6$.

4. Clean the entries. It is possible that one rich cluster can be
recognized two or more times in the above procedures when targeting
different bright galaxies, and hence some entries should be merged to
one cluster if their redshift difference is less than 0.1 and a
projected distance is less than $1.5\,r_{500}$.

We find 79,498 clusters through the above procedures, of which 54,079
can match the clusters in the WHL12 catalog with a redshift difference
of 0.1 and a projected distance of $1.5\,r_{500}$. The remaining
25,419 are newly identified clusters, as listed in
Table~\ref{tab3}. Most of them have a high redshift of $z>0.4$ as
shown in Figure~\ref{histzall}.

If we combine the cluster sample in the updated WHL12 catalog and the
complementary catalog of newly identified 25,419 clusters, we get
158,103 clusters, of which 96,257 have a richness of
$R_{L\ast,500}\ge20$. Figure~\ref{sky} shows the sky coverage and
radial distribution in a sky slice for clusters in the combined
catalog. In the SDSS spectroscopic survey region, 89\% of clusters
have spectroscopic redshifts.

\section{Conclusions}

In this paper, we calibrated the optical mass proxy of galaxy clusters
and applied it to update the WHL12 cluster catalog.

We collected clusters with known masses ($M_{500}$) derived from X-ray
or SZ measurements from the literature, and rescaled them for
consistency. The composite mass calibration sample consists of 1191
clusters in the redshift range $0.05<z<0.75$ with mass
$M_{500}>0.3\times10^{14}~M_{\odot}$, and all of them are limited to
the SDSS sky coverage to obtain the optical spectroscopic and
photometric data.

We established the scaling relations between the total $r$-band
optical luminosity of member galaxies $L_{\rm 1Mpc}$ within 1~Mpc and
the cluster radius ($r_{500}$). After correcting for weak redshift
dependence, the scaling relation can be used to estimate the cluster
radius from optical data.
The cluster mass $M_{500}$ is found to be well-correlated with the
total $r$-band luminosity $L_{500}$, which evolves significantly 
with redshift. Considering the redshift evolution, we defined the
new cluster richness as $R_{L\ast,500}=L_{500}\,E(z)^{1.40}$, which can
be used as a mass proxy independent of redshift and is valid in
the redshift range of $0.05<z<0.75$. The scatter of mass estimates
is about 0.17 dex.

We updated the spectroscopic redshifts for the WHL12 clusters using
the SDSS DR12 spectroscopic data, and then obtain their newly defined
richness.  We identify 25,419 complementary clusters at high
redshifts around bright galaxies with a spectroscopic redshift in the
SDSS DR12. The finial combined catalog contains 158,103 clusters in
total, and about 89\% of them have spectroscopic redshifts in the SDSS
spectroscopic survey region.

\acknowledgments

We thank the referee for helpful comments. The authors are supported
by the National Natural Science Foundation (NNSF) of China (11103032
and 11473034) and the Young Researcher Grant of National Astronomical
Observatories, Chinese Academy of Sciences.
Funding for SDSS-III has been provided by theAlfred P. Sloan
Foundation, the Participating Institutions, the National Science
Foundation, and theUSDepartment of Energy.  The SDSS-III Web site is
http://www.sdss3.org/.
SDSS-III is managed by the Astrophysical Research Consortium for the
Participating Institutions of the SDSS-III Collaboration including the
University of Arizona, the Brazilian Participation Group, Brookhaven
National Laboratory, University of Cambridge, University of Florida,
the French Participation Group, the German Participation Group, the
Instituto de Astrofisica de Canarias, the Michigan State/Notre
Dame/JINA Participation Group, Johns Hopkins University, Lawrence
Berkeley National Laboratory, Max Planck Institute for Astrophysics,
NewMexico State University, New York University, Ohio StateUniversity,
Pennsylvania State University,University of Portsmouth, Princeton
University, the Spanish Participation Group, University of Tokyo,
University of Utah,Vanderbilt University, University of Virginia,
University of Washington, and Yale University.


\begin{thebibliography}{82}
\expandafter\ifx\csname natexlab\endcsname\relax\def\natexlab#1{#1}\fi

\bibitem[{{Alam} {et~al.}(2015){Alam}, {Albareti}, {Allende Prieto}, {Anders},
  {Anderson}, {Andrews}, {Armengaud}, {Aubourg}, {Bailey}, {Bautista}, \&
  et~al.}]{dr12+15}
{Alam}, S., {Albareti}, F.~D., {Allende Prieto}, C., {et~al.} 2015, arXiv:1501.00963

\bibitem[{{Allen} {et~al.}(2011){Allen}, {Evrard}, \& {Mantz}}]{aem11}
{Allen}, S.~W., {Evrard}, A.~E., \& {Mantz}, A.~B. 2011, \araa, 49, 409

\bibitem[{{Andreon} \& {Congdon}(2014)}]{ac14}
{Andreon}, S., \& {Congdon}, P. 2014, \aap, 568, A23

\bibitem[{{Arnaud} {et~al.}(2010){Arnaud}, {Pratt}, {Piffaretti},
  {B{\"o}hringer}, {Croston}, \& {Pointecouteau}}]{app+10}
{Arnaud}, M., {Pratt}, G.~W., {Piffaretti}, R., {et~al.} 2010, \aap, 517, A92

\bibitem[{{Bardeau} {et~al.}(2007){Bardeau}, {Soucail}, {Kneib}, {Czoske},
  {Ebeling}, {Hudelot}, {Smail}, \& {Smith}}]{bsk+07}
{Bardeau}, S., {Soucail}, G., {Kneib}, J.-P., {et~al.} 2007, \aap, 470, 449

\bibitem[{{Berlind} {et~al.}(2006){Berlind}, {Frieman}, {Weinberg}, {Blanton},
  {Warren}, {Abazajian}, {Scranton}, {Hogg}, {Scoccimarro}, {Bahcall},
  {Brinkmann}, {Gott}, {Kleinman}, {Krzesinski}, {Lee}, {Miller}, {Nitta},
  {Schneider}, {Tucker}, \& {Zehavi}}]{bfw+06}
{Berlind}, A.~A., {Frieman}, J., {Weinberg}, D.~H., {et~al.} 2006, \apjs, 167,
  1

\bibitem[{{Bird}(1995)}]{bir95}
{Bird}, C.~M. 1995, \apjl, 445, L81

\bibitem[{{Blanton} {et~al.}(2003){Blanton}, {Hogg}, {Bahcall}, {Brinkmann},
  {Britton}, {Connolly}, {Csabai}, {Fukugita}, {Loveday}, {Meiksin}, {Munn},
  {Nichol}, {Okamura}, {Quinn}, {Schneider}, {Shimasaku}, {Strauss}, {Tegmark},
  {Vogeley}, \& {Weinberg}}]{bhb+03}
{Blanton}, M.~R., {Hogg}, D.~W., {Bahcall}, N.~A., {et~al.} 2003, \apj, 592,
  819

\bibitem[{{Bode} {et~al.}(2012){Bode}, {Ostriker}, {Cen}, \& {Trac}}]{boc+12}
{Bode}, P., {Ostriker}, J.~P., {Cen}, R., \& {Trac}, H. 2012, arXiv:1204.1762

\bibitem[{{B{\"o}hringer} {et~al.}(2000){B{\"o}hringer}, {Voges}, {Huchra},
  {McLean}, {Giacconi}, {Rosati}, {Burg}, {Mader}, {Schuecker}, {Simi{\c c}},
  {Komossa}, {Reiprich}, {Retzlaff}, \& {Tr{\"u}mper}}]{bvh+00}
{B{\"o}hringer}, H., {Voges}, W., {Huchra}, J.~P., {et~al.} 2000, \apjs, 129,
  435

\bibitem[{{B{\"o}hringer} {et~al.}(2004){B{\"o}hringer}, {Schuecker}, {Guzzo},
  {Collins}, {Voges}, {Cruddace}, {Ortiz-Gil}, {Chincarini}, {De Grandi},
  {Edge}, {MacGillivray}, {Neumann}, {Schindler}, \& {Shaver}}]{bsg+04}
{B{\"o}hringer}, H., {Schuecker}, P., {Guzzo}, L., {et~al.} 2004, \aap, 425,
  367

\bibitem[{{Brunetti} {et~al.}(2009){Brunetti}, {Cassano}, {Dolag}, \&
  {Setti}}]{bcd+09}
{Brunetti}, G., {Cassano}, R., {Dolag}, K., \& {Setti}, G. 2009, \aap, 507, 661

\bibitem[{{Bruzual} \& {Charlot}(2003)}]{bc03}
{Bruzual}, G., \& {Charlot}, S. 2003, \mnras, 344, 1000

\bibitem[{{Burenin} {et~al.}(2007){Burenin}, {Vikhlinin}, {Hornstrup},
  {Ebeling}, {Quintana}, \& {Mescheryakov}}]{bvh+07}
{Burenin}, R.~A., {Vikhlinin}, A., {Hornstrup}, A., {et~al.} 2007, \apjs, 172,
  561

\bibitem[{{Carlstrom} {et~al.}(2002){Carlstrom}, {Holder}, \& {Reese}}]{che02}
{Carlstrom}, J.~E., {Holder}, G.~P., \& {Reese}, E.~D. 2002, \araa, 40, 643

\bibitem[{{Chabrier}(2003)}]{cha03}
{Chabrier}, G. 2003, \pasp, 115, 763

\bibitem[{{Crawford} {et~al.}(1999){Crawford}, {Allen}, {Ebeling}, {Edge}, \&
  {Fabian}}]{cae+99}
{Crawford}, C.~S., {Allen}, S.~W., {Ebeling}, H., {Edge}, A.~C., \& {Fabian},
  A.~C. 1999, \mnras, 306, 857

\bibitem[{{Crawford} {et~al.}(2009){Crawford}, {Bershady}, \&
  {Hoessel}}]{cbh09}
{Crawford}, S.~M., {Bershady}, M.~A., \& {Hoessel}, J.~G. 2009, \apj, 690, 1158

\bibitem[{{Cruddace} {et~al.}(2002){Cruddace}, {Voges}, {B{\"o}hringer},
  {Collins}, {Romer}, {MacGillivray}, {Yentis}, {Schuecker}, {Ebeling}, \& {De
  Grandi}}]{cvb+02}
{Cruddace}, R., {Voges}, W., {B{\"o}hringer}, H., {et~al.} 2002, \apjs, 140,
  239

\bibitem[{{Csabai} {et~al.}(2007){Csabai}, {Dobos}, {Trencs{\'e}ni},
  {Herczegh}, {J{\'o}zsa}, {Purger}, {Budav{\'a}ri}, \& {Szalay}}]{cdt+07}
{Csabai}, I., {Dobos}, L., {Trencs{\'e}ni}, M., {et~al.} 2007, Astronomische
  Nachrichten, 328, 852

\bibitem[{{Dahle}(2006)}]{dah06}
{Dahle}, H. 2006, \apj, 653, 954

\bibitem[{{Dawson} {et~al.}(2013){Dawson}, {Schlegel}, {Ahn}, {Anderson},
  {Aubourg}, {Bailey}, {Barkhouser}, {Bautista}, {Beifiori}, {Berlind},
  {Bhardwaj}, {Bizyaev}, {Blake}, {Blanton}, {Blomqvist}, {Bolton}, {Borde},
  {Bovy}, {Brandt}, {Brewington}, {Brinkmann}, {Brown}, {Brownstein}, {Bundy},
  {Busca}, {Carithers}, {Carnero}, {Carr}, {Chen}, {Comparat}, {Connolly},
  {Cope}, {Croft}, {Cuesta}, {da Costa}, {Davenport}, {Delubac}, {de Putter},
  {Dhital}, {Ealet}, {Ebelke}, {Eisenstein}, {Escoffier}, {Fan}, {Filiz Ak},
  {Finley}, {Font-Ribera}, {G{\'e}nova-Santos}, {Gunn}, {Guo}, {Haggard},
  {Hall}, {Hamilton}, {Harris}, {Harris}, {Ho}, {Hogg}, {Holder}, {Honscheid},
  {Huehnerhoff}, {Jordan}, {Jordan}, {Kauffmann}, {Kazin}, {Kirkby}, {Klaene},
  {Kneib}, {Le Goff}, {Lee}, {Long}, {Loomis}, {Lundgren}, {Lupton}, {Maia},
  {Makler}, {Malanushenko}, {Malanushenko}, {Mandelbaum}, {Manera}, {Maraston},
  {Margala}, {Masters}, {McBride}, {McDonald}, {McGreer}, {McMahon}, {Mena},
  {Miralda-Escud{\'e}}, {Montero-Dorta}, {Montesano}, {Muna}, {Myers},
  {Naugle}, {Nichol}, {Noterdaeme}, {Nuza}, {Olmstead}, {Oravetz}, {Oravetz},
  {Owen}, {Padmanabhan}, {Palanque-Delabrouille}, {Pan}, {Parejko},
  {P{\^a}ris}, {Percival}, {P{\'e}rez-Fournon}, {P{\'e}rez-R{\`a}fols},
  {Petitjean}, {Pfaffenberger}, {Pforr}, {Pieri}, {Prada}, {Price-Whelan},
  {Raddick}, {Rebolo}, {Rich}, {Richards}, {Rockosi}, {Roe}, {Ross}, {Ross},
  {Rossi}, {Rubi{\~n}o-Martin}, {Samushia}, {S{\'a}nchez}, {Sayres}, {Schmidt},
  {Schneider}, {Sc{\'o}ccola}, {Seo}, {Shelden}, {Sheldon}, {Shen}, {Shu},
  {Slosar}, {Smee}, {Snedden}, {Stauffer}, {Steele}, {Strauss}, {Streblyanska},
  {Suzuki}, {Swanson}, {Tal}, {Tanaka}, {Thomas}, {Tinker}, {Tojeiro},
  {Tremonti}, {Vargas Maga{\~n}a}, {Verde}, {Viel}, {Wake}, {Watson}, {Weaver},
  {Weinberg}, {Weiner}, {West}, {White}, {Wood-Vasey}, {Yeche}, {Zehavi},
  {Zhao}, \& {Zheng}}]{dsa+13}
{Dawson}, K.~S., {Schlegel}, D.~J., {Ahn}, C.~P., {et~al.} 2013, \aj, 145, 10

\bibitem[{{Ebeling} {et~al.}(1998){Ebeling}, {Edge}, {Bohringer}, {Allen},
  {Crawford}, {Fabian}, {Voges}, \& {Huchra}}]{eeb+98}
{Ebeling}, H., {Edge}, A.~C., {Bohringer}, H., {et~al.} 1998, \mnras, 301, 881

\bibitem[{{Ebeling} {et~al.}(2001){Ebeling}, {Edge}, \& {Henry}}]{eeh01}
{Ebeling}, H., {Edge}, A.~C., \& {Henry}, J.~P. 2001, \apj, 553, 668

\bibitem[{{Ebeling} {et~al.}(2002){Ebeling}, {Mullis}, \& {Tully}}]{emt02}
{Ebeling}, H., {Mullis}, C.~R., \& {Tully}, R.~B. 2002, \apj, 580, 774

\bibitem[{{Eisenstein} {et~al.}(2001){Eisenstein}, {Annis}, {Gunn}, {Szalay},
  {Connolly}, {Nichol}, {Bahcall}, {Bernardi}, {Burles}, {Castander},
  {Fukugita}, {Hogg}, {Ivezi{\'c}}, {Knapp}, {Lupton}, {Narayanan}, {Postman},
  {Reichart}, {Richmond}, {Schneider}, {Schlegel}, {Strauss}, {SubbaRao},
  {Tucker}, {Vanden Berk}, {Vogeley}, {Weinberg}, \& {Yanny}}]{eag+01}
{Eisenstein}, D.~J., {Annis}, J., {Gunn}, J.~E., {et~al.} 2001, \aj, 122, 2267

\bibitem[{{Finoguenov} {et~al.}(2001){Finoguenov}, {Reiprich}, \&
  {B{\"o}hringer}}]{frb01}
{Finoguenov}, A., {Reiprich}, T.~H., \& {B{\"o}hringer}, H. 2001, \aap, 368,
  749

\bibitem[{{Gioia} {et~al.}(1990){Gioia}, {Maccacaro}, {Schild}, {Wolter},
  {Stocke}, {Morris}, \& {Henry}}]{gms+90}
{Gioia}, I.~M., {Maccacaro}, T., {Schild}, R.~E., {et~al.} 1990, \apjs, 72, 567

\bibitem[{{Girardi} {et~al.}(1998){Girardi}, {Giuricin}, {Mardirossian},
  {Mezzetti}, \& {Boschin}}]{ggm+98}
{Girardi}, M., {Giuricin}, G., {Mardirossian}, F., {Mezzetti}, M., \&
  {Boschin}, W. 1998, \apj, 505, 74

\bibitem[{{Gladders} \& {Yee}(2000)}]{gy00}
{Gladders}, M.~D., \& {Yee}, H.~K.~C. 2000, \aj, 120, 2148

\bibitem[{{Gu} {et~al.}(2013){Gu}, {Gandhi}, {Inada}, {Kawaharada}, {Kodama},
  {Konami}, {Nakazawa}, {Shimasaku}, {Xu}, \& {Makishima}}]{ggi+13}
{Gu}, L., {Gandhi}, P., {Inada}, N., {et~al.} 2013, \apj, 767, 157

\bibitem[{{Hao} {et~al.}(2010){Hao}, {McKay}, {Koester}, {Rykoff}, {Rozo},
  {Annis}, {Wechsler}, {Evrard}, {Siegel}, {Becker}, {Busha}, {Gerdes},
  {Johnston}, \& {Sheldon}}]{hmk+10}
{Hao}, J., {McKay}, T.~A., {Koester}, B.~P., {et~al.} 2010, \apjs, 191, 254

\bibitem[{{Hasselfield} {et~al.}(2013){Hasselfield}, {Hilton}, {Marriage},
  {Addison}, {Barrientos}, {Battaglia}, {Battistelli}, {Bond}, {Crichton},
  {Das}, {Devlin}, {Dicker}, {Dunkley}, {D{\"u}nner}, {Fowler}, {Gralla},
  {Hajian}, {Halpern}, {Hincks}, {Hlozek}, {Hughes}, {Infante}, {Irwin},
  {Kosowsky}, {Marsden}, {Menanteau}, {Moodley}, {Niemack}, {Nolta}, {Page},
  {Partridge}, {Reese}, {Schmitt}, {Sehgal}, {Sherwin}, {Sievers}, {Sif{\'o}n},
  {Spergel}, {Staggs}, {Swetz}, {Switzer}, {Thornton}, {Trac}, \&
  {Wollack}}]{hhm+13}
{Hasselfield}, M., {Hilton}, M., {Marriage}, T.~A., {et~al.} 2013, \jcap, 7, 8

\bibitem[{{Henry} {et~al.}(2006){Henry}, {Mullis}, {Voges}, {B{\"o}hringer},
  {Briel}, {Gioia}, \& {Huchra}}]{hmv+06}
{Henry}, J.~P., {Mullis}, C.~R., {Voges}, W., {et~al.} 2006, \apjs, 162, 304

\bibitem[{{Horner} {et~al.}(2008){Horner}, {Perlman}, {Ebeling}, {Jones},
  {Scharf}, {Wegner}, {Malkan}, \& {Maughan}}]{hpe+08}
{Horner}, D.~J., {Perlman}, E.~S., {Ebeling}, H., {et~al.} 2008, \apjs, 176,
  374

\bibitem[{{Koester} {et~al.}(2007){Koester}, {McKay}, {Annis}, {Wechsler},
  {Evrard}, {Bleem}, {Becker}, {Johnston}, {Sheldon}, {Nichol}, {Miller},
  {Scranton}, {Bahcall}, {Barentine}, {Brewington}, {Brinkmann}, {Harvanek},
  {Kleinman}, {Krzesinski}, {Long}, {Nitta}, {Schneider}, {Sneddin}, {Voges},
  \& {York}}]{kma+07}
{Koester}, B.~P., {McKay}, T.~A., {Annis}, J., {et~al.} 2007, \apj, 660, 239

\bibitem[{{Kravtsov} {et~al.}(2006){Kravtsov}, {Vikhlinin}, \& {Nagai}}]{kvn06}
{Kravtsov}, A.~V., {Vikhlinin}, A., \& {Nagai}, D. 2006, \apj, 650, 128

\bibitem[{{Lin} {et~al.}(2006){Lin}, {Mohr}, {Gonzalez}, \&
  {Stanford}}]{lmg+06}
{Lin}, Y.-T., {Mohr}, J.~J., {Gonzalez}, A.~H., \& {Stanford}, S.~A. 2006,
  \apjl, 650, L99

\bibitem[{{Lin} {et~al.}(2004){Lin}, {Mohr}, \& {Stanford}}]{lms04}
{Lin}, Y.-T., {Mohr}, J.~J., \& {Stanford}, S.~A. 2004, \apj, 610, 745

\bibitem[{{Lupton} {et~al.}(2001){Lupton}, {Gunn}, {Ivezi{\'c}}, {Knapp}, \&
  {Kent}}]{lgi+01}
{Lupton}, R., {Gunn}, J.~E., {Ivezi{\'c}}, Z., {Knapp}, G.~R., \& {Kent}, S.
  2001, in Astronomical Society of the Pacific Conference Series, Vol. 238,
  Astronomical Data Analysis Software and Systems X, ed. F.~R. {Harnden}, Jr.,
  F.~A. {Primini}, \& H.~E. {Payne}, 269

\bibitem[{{Mantz} {et~al.}(2010){Mantz}, {Allen}, {Ebeling}, {Rapetti}, \&
  {Drlica-Wagner}}]{mae+10}
{Mantz}, A., {Allen}, S.~W., {Ebeling}, H., {Rapetti}, D., \& {Drlica-Wagner},
  A. 2010, \mnras, 406, 1773

\bibitem[{{Merch{\'a}n} \& {Zandivarez}(2005)}]{mz05}
{Merch{\'a}n}, M.~E., \& {Zandivarez}, A. 2005, \apj, 630, 759

\bibitem[{{Mullis} {et~al.}(2003){Mullis}, {McNamara}, {Quintana}, {Vikhlinin},
  {Henry}, {Gioia}, {Hornstrup}, {Forman}, \& {Jones}}]{mmq+03}
{Mullis}, C.~R., {McNamara}, B.~R., {Quintana}, H., {et~al.} 2003, \apj, 594,
  154

\bibitem[{{Oguri}(2014)}]{ogu14}
{Oguri}, M. 2014, \mnras, 444, 147

\bibitem[{{Okabe} {et~al.}(2010){Okabe}, {Takada}, {Umetsu}, {Futamase}, \&
  {Smith}}]{otu+10}
{Okabe}, N., {Takada}, M., {Umetsu}, K., {Futamase}, T., \& {Smith}, G.~P.
  2010, \pasj, 62, 811

\bibitem[{{Piffaretti} {et~al.}(2011){Piffaretti}, {Arnaud}, {Pratt},
  {Pointecouteau}, \& {Melin}}]{pap+11}
{Piffaretti}, R., {Arnaud}, M., {Pratt}, G.~W., {Pointecouteau}, E., \&
  {Melin}, J.-B. 2011, \aap, 534, A109

\bibitem[{{Pipino} {et~al.}(2011){Pipino}, {Szabo}, {Pierpaoli}, {MacKenzie},
  \& {Dong}}]{psp+11}
{Pipino}, A., {Szabo}, T., {Pierpaoli}, E., {MacKenzie}, S.~M., \& {Dong}, F.
  2011, \mnras, 417, 2817

\bibitem[{{Planck Collaboration} {et~al.}(2014){Planck Collaboration}, {Ade},
  {Aghanim}, {Armitage-Caplan}, {Arnaud}, {Ashdown}, {Atrio-Barandela},
  {Aumont}, {Baccigalupi}, {Banday}, \& et~al.}]{planck14}
{Planck Collaboration}, {Ade}, P.~A.~R., {Aghanim}, N., {et~al.} 2014, \aap,
  571, A20

\bibitem[{{Planck Collaboration} {et~al.}(2015){Planck Collaboration}, {Ade},
  {Aghanim}, {Arnaud}, {Ashdown}, {Aumont}, {Baccigalupi}, {Banday},
  {Barreiro}, {Barrena}, \& et~al.}]{planck15}
---. 2015, arXiv:1502.01598

\bibitem[{{Popesso} {et~al.}(2007){Popesso}, {Biviano}, {B{\"o}hringer}, \&
  {Romaniello}}]{pbb+07}
{Popesso}, P., {Biviano}, A., {B{\"o}hringer}, H., \& {Romaniello}, M. 2007,
  \aap, 464, 451

\bibitem[{{Popesso} {et~al.}(2005){Popesso}, {Biviano}, {B{\"o}hringer},
  {Romaniello}, \& {Voges}}]{pbb+05}
{Popesso}, P., {Biviano}, A., {B{\"o}hringer}, H., {Romaniello}, M., \&
  {Voges}, W. 2005, \aap, 433, 431

\bibitem[{{Popesso} {et~al.}(2004){Popesso}, {B{\"o}hringer}, {Brinkmann},
  {Voges}, \& {York}}]{pbb+04}
{Popesso}, P., {B{\"o}hringer}, H., {Brinkmann}, J., {Voges}, W., \& {York},
  D.~G. 2004, \aap, 423, 449

\bibitem[{{Pratt} {et~al.}(2009){Pratt}, {Croston}, {Arnaud}, \&
  {B{\"o}hringer}}]{pca+09}
{Pratt}, G.~W., {Croston}, J.~H., {Arnaud}, M., \& {B{\"o}hringer}, H. 2009,
  \aap, 498, 361

\bibitem[{{Reichardt} {et~al.}(2013){Reichardt}, {Stalder}, {Bleem}, {Montroy},
  {Aird}, {Andersson}, {Armstrong}, {Ashby}, {Bautz}, {Bayliss}, {Bazin},
  {Benson}, {Brodwin}, {Carlstrom}, {Chang}, {Cho}, {Clocchiatti}, {Crawford},
  {Crites}, {de Haan}, {Desai}, {Dobbs}, {Dudley}, {Foley}, {Forman}, {George},
  {Gladders}, {Gonzalez}, {Halverson}, {Harrington}, {High}, {Holder},
  {Holzapfel}, {Hoover}, {Hrubes}, {Jones}, {Joy}, {Keisler}, {Knox}, {Lee},
  {Leitch}, {Liu}, {Lueker}, {Luong-Van}, {Mantz}, {Marrone}, {McDonald},
  {McMahon}, {Mehl}, {Meyer}, {Mocanu}, {Mohr}, {Murray}, {Natoli}, {Padin},
  {Plagge}, {Pryke}, {Rest}, {Ruel}, {Ruhl}, {Saliwanchik}, {Saro}, {Sayre},
  {Schaffer}, {Shaw}, {Shirokoff}, {Song}, {Spieler}, {Staniszewski}, {Stark},
  {Story}, {Stubbs}, {{\v S}uhada}, {van Engelen}, {Vanderlinde}, {Vieira},
  {Vikhlinin}, {Williamson}, {Zahn}, \& {Zenteno}}]{rsb+13}
{Reichardt}, C.~L., {Stalder}, B., {Bleem}, L.~E., {et~al.} 2013, \apj, 763,
  127

\bibitem[{{Reiprich} \& {B{\"o}hringer}(2002)}]{rb02}
{Reiprich}, T.~H., \& {B{\"o}hringer}, H. 2002, \apj, 567, 716

\bibitem[{{Romer} {et~al.}(2000){Romer}, {Nichol}, {Holden}, {Ulmer}, {Pildis},
  {Merrelli}, {Adami}, {Burke}, {Collins}, {Metevier}, {Kron}, \&
  {Commons}}]{rnh+00}
{Romer}, A.~K., {Nichol}, R.~C., {Holden}, B.~P., {et~al.} 2000, \apjs, 126,
  209

\bibitem[{{Rozo} {et~al.}(2009){Rozo}, {Rykoff}, {Koester}, {McKay}, {Hao},
  {Evrard}, {Wechsler}, {Hansen}, {Sheldon}, {Johnston}, {Becker}, {Annis},
  {Bleem}, \& {Scranton}}]{rrk+09}
{Rozo}, E., {Rykoff}, E.~S., {Koester}, B.~P., {et~al.} 2009, \apj, 703, 601

\bibitem[{{Rykoff} {et~al.}(2012){Rykoff}, {Koester}, {Rozo}, {Annis},
  {Evrard}, {Hansen}, {Hao}, {Johnston}, {McKay}, \& {Wechsler}}]{rkr+12}
{Rykoff}, E.~S., {Koester}, B.~P., {Rozo}, E., {et~al.} 2012, \apj, 746, 178

\bibitem[{{Rykoff} {et~al.}(2014){Rykoff}, {Rozo}, {Busha}, {Cunha},
  {Finoguenov}, {Evrard}, {Hao}, {Koester}, {Leauthaud}, {Nord}, {Pierre},
  {Reddick}, {Sadibekova}, {Sheldon}, \& {Wechsler}}]{rrb+14}
{Rykoff}, E.~S., {Rozo}, E., {Busha}, M.~T., {et~al.} 2014, \apj, 785, 104

\bibitem[{{Sif{\'o}n} {et~al.}(2013){Sif{\'o}n}, {Menanteau}, {Hasselfield},
  {Marriage}, {Hughes}, {Barrientos}, {Gonz{\'a}lez}, {Infante}, {Addison},
  {Baker}, {Battaglia}, {Bond}, {Crichton}, {Das}, {Devlin}, {Dunkley},
  {D{\"u}nner}, {Gralla}, {Hajian}, {Hilton}, {Hincks}, {Kosowsky}, {Marsden},
  {Moodley}, {Niemack}, {Nolta}, {Page}, {Partridge}, {Reese}, {Sehgal},
  {Sievers}, {Spergel}, {Staggs}, {Thornton}, {Trac}, \& {Wollack}}]{smh+13}
{Sif{\'o}n}, C., {Menanteau}, F., {Hasselfield}, M., {et~al.} 2013, \apj, 772,
  25

\bibitem[{{Stoughton} {et~al.}(2002){Stoughton}, {Lupton}, {Bernardi},
  {Blanton}, {Burles}, {Castander}, {Connolly}, {Eisenstein}, {Frieman},
  {Hennessy}, {Hindsley}, {Ivezi{\'c}}, {Kent}, {Kunszt}, {Lee}, {Meiksin},
  {Munn}, {Newberg}, {Nichol}, {Nicinski}, {Pier}, {Richards}, {Richmond},
  {Schlegel}, {Smith}, {Strauss}, {SubbaRao}, {Szalay}, {Thakar}, {Tucker},
  {Vanden Berk}, {Yanny}, {Adelman}, {Anderson}, {Anderson}, {Annis},
  {Bahcall}, {Bakken}, {Bartelmann}, {Bastian}, {Bauer}, {Berman},
  {B{\"o}hringer}, {Boroski}, {Bracker}, {Briegel}, {Briggs}, {Brinkmann},
  {Brunner}, {Carey}, {Carr}, {Chen}, {Christian}, {Colestock}, {Crocker},
  {Csabai}, {Czarapata}, {Dalcanton}, {Davidsen}, {Davis}, {Dehnen},
  {Dodelson}, {Doi}, {Dombeck}, {Donahue}, {Ellman}, {Elms}, {Evans}, {Eyer},
  {Fan}, {Federwitz}, {Friedman}, {Fukugita}, {Gal}, {Gillespie}, {Glazebrook},
  {Gray}, {Grebel}, {Greenawalt}, {Greene}, {Gunn}, {de Haas}, {Haiman},
  {Haldeman}, {Hall}, {Hamabe}, {Hansen}, {Harris}, {Harris}, {Harvanek},
  {Hawley}, {Hayes}, {Heckman}, {Helmi}, {Henden}, {Hogan}, {Hogg}, {Holmgren},
  {Holtzman}, {Huang}, {Hull}, {Ichikawa}, {Ichikawa}, {Johnston}, {Kauffmann},
  {Kim}, {Kimball}, {Kinney}, {Klaene}, {Kleinman}, {Klypin}, {Knapp},
  {Korienek}, {Krolik}, {Kron}, {Krzesi{\'n}ski}, {Lamb}, {Leger},
  {Limmongkol}, {Lindenmeyer}, {Long}, {Loomis}, {Loveday}, {MacKinnon},
  {Mannery}, {Mantsch}, {Margon}, {McGehee}, {McKay}, {McLean}, {Menou},
  {Merelli}, {Mo}, {Monet}, {Nakamura}, {Narayanan}, {Nash}, {Neilsen},
  {Newman}, {Nitta}, {Odenkirchen}, {Okada}, {Okamura}, {Ostriker}, {Owen},
  {Pauls}, {Peoples}, {Peterson}, {Petravick}, {Pope}, {Pordes}, {Postman},
  {Prosapio}, {Quinn}, {Rechenmacher}, {Rivetta}, {Rix}, {Rockosi}, {Rosner},
  {Ruthmansdorfer}, {Sandford}, {Schneider}, {Scranton}, {Sekiguchi}, {Sergey},
  {Sheth}, {Shimasaku}, {Smee}, {Snedden}, {Stebbins}, {Stubbs}, {Szapudi},
  {Szkody}, {Szokoly}, {Tabachnik}, {Tsvetanov}, {Uomoto}, {Vogeley}, {Voges},
  {Waddell}, {Walterbos}, {Wang}, {Watanabe}, {Weinberg}, {White}, {White},
  {Wilhite}, {Wolfe}, {Yasuda}, {York}, {Zehavi}, \& {Zheng}}]{slb+02}
{Stoughton}, C., {Lupton}, R.~H., {Bernardi}, M., {et~al.} 2002, \aj, 123, 485

\bibitem[{{Strauss} {et~al.}(2002){Strauss}, {Weinberg}, {Lupton}, {Narayanan},
  {Annis}, {Bernardi}, {Blanton}, {Burles}, {Connolly}, {Dalcanton}, {Doi},
  {Eisenstein}, {Frieman}, {Fukugita}, {Gunn}, {Ivezi{\'c}}, {Kent}, {Kim},
  {Knapp}, {Kron}, {Munn}, {Newberg}, {Nichol}, {Okamura}, {Quinn}, {Richmond},
  {Schlegel}, {Shimasaku}, {SubbaRao}, {Szalay}, {Vanden Berk}, {Vogeley},
  {Yanny}, {Yasuda}, {York}, \& {Zehavi}}]{swl+02}
{Strauss}, M.~A., {Weinberg}, D.~H., {Lupton}, R.~H., {et~al.} 2002, \aj, 124,
  1810

\bibitem[{{Szabo} {et~al.}(2011){Szabo}, {Pierpaoli}, {Dong}, {Pipino}, \&
  {Gunn}}]{spd+11}
{Szabo}, T., {Pierpaoli}, E., {Dong}, F., {Pipino}, A., \& {Gunn}, J. 2011,
  \apj, 736, 21

\bibitem[{{Takey} {et~al.}(2013){Takey}, {Schwope}, \& {Lamer}}]{tsl13}
{Takey}, A., {Schwope}, A., \& {Lamer}, G. 2013, \aap, 558, A75

\bibitem[{{Takey} {et~al.}(2014){Takey}, {Schwope}, \& {Lamer}}]{tsl14}
---. 2014, \aap, 564, A54

\bibitem[{{Tempel} {et~al.}(2014){Tempel}, {Tamm}, {Gramann}, {Tuvikene},
  {Liivam{\"a}gi}, {Suhhonenko}, {Kipper}, {Einasto}, \& {Saar}}]{ttg+14}
{Tempel}, E., {Tamm}, A., {Gramann}, M., {et~al.} 2014, \aap, 566, A1

\bibitem[{{Trac} {et~al.}(2011){Trac}, {Bode}, \& {Ostriker}}]{tbo11}
{Trac}, H., {Bode}, P., \& {Ostriker}, J.~P. 2011, \apj, 727, 94

\bibitem[{{Vikhlinin} {et~al.}(2006){Vikhlinin}, {Kravtsov}, {Forman}, {Jones},
  {Markevitch}, {Murray}, \& {Van Speybroeck}}]{vkf+06}
{Vikhlinin}, A., {Kravtsov}, A., {Forman}, W., {et~al.} 2006, \apj, 640, 691

\bibitem[{{Vikhlinin} {et~al.}(2009{\natexlab{a}}){Vikhlinin}, {Burenin},
  {Ebeling}, {Forman}, {Hornstrup}, {Jones}, {Kravtsov}, {Murray}, {Nagai},
  {Quintana}, \& {Voevodkin}}]{vbe+09}
{Vikhlinin}, A., {Burenin}, R.~A., {Ebeling}, H., {et~al.} 2009{\natexlab{a}},
  \apj, 692, 1033

\bibitem[{{Vikhlinin} {et~al.}(2009{\natexlab{b}}){Vikhlinin}, {Kravtsov},
  {Burenin}, {Ebeling}, {Forman}, {Hornstrup}, {Jones}, {Murray}, {Nagai},
  {Quintana}, \& {Voevodkin}}]{vkb+09}
{Vikhlinin}, A., {Kravtsov}, A.~V., {Burenin}, R.~A., {et~al.}
  2009{\natexlab{b}}, \apj, 692, 1060

\bibitem[{{Voit}(2005)}]{voi05}
{Voit}, G.~M. 2005, Reviews of Modern Physics, 77, 207

\bibitem[{{von der Linden} {et~al.}(2014{\natexlab{a}}){von der Linden},
  {Mantz}, {Allen}, {Applegate}, {Kelly}, {Morris}, {Wright}, {Allen},
  {Burchat}, {Burke}, {Donovan}, \& {Ebeling}}]{vma+14}
{von der Linden}, A., {Mantz}, A., {Allen}, S.~W., {et~al.} 2014{\natexlab{a}},
  \mnras, 443, 1973

\bibitem[{{von der Linden} {et~al.}(2014{\natexlab{b}}){von der Linden},
  {Allen}, {Applegate}, {Kelly}, {Allen}, {Ebeling}, {Burchat}, {Burke},
  {Donovan}, {Morris}, {Blandford}, {Erben}, \& {Mantz}}]{oaa+14}
{von der Linden}, A., {Allen}, M.~T., {Applegate}, D.~E., {et~al.}
  2014{\natexlab{b}}, \mnras, 439, 2

\bibitem[{{Wen} {et~al.}(2009){Wen}, {Han}, \& {Liu}}]{whl09}
{Wen}, Z.~L., {Han}, J.~L., \& {Liu}, F.~S. 2009, \apjs, 183, 197

\bibitem[{{Wen} {et~al.}(2010){Wen}, {Han}, \& {Liu}}]{whl10}
---. 2010, \mnras, 407, 533

\bibitem[{{Wen} {et~al.}(2012){Wen}, {Han}, \& {Liu}}]{whl12}
---. 2012, \apjs, 199, 34

\bibitem[{{Wetzel} {et~al.}(2012){Wetzel}, {Tinker}, \& {Conroy}}]{wtc12}
{Wetzel}, A.~R., {Tinker}, J.~L., \& {Conroy}, C. 2012, \mnras, 424, 232

\bibitem[{{Wiesner} {et~al.}(2015){Wiesner}, {Lin}, \& {Soares-Santos}}]{wls15}
{Wiesner}, M.~P., {Lin}, H., \& {Soares-Santos}, M. 2015, arXiv:1501.06893

\bibitem[{{Yang} {et~al.}(2007){Yang}, {Mo}, {van den Bosch}, {Pasquali}, {Li},
  \& {Barden}}]{ymv+07}
{Yang}, X., {Mo}, H.~J., {van den Bosch}, F.~C., {et~al.} 2007, \apj, 671, 153

\bibitem[{{York} {et~al.}(2000){York}, {Adelman}, {Anderson}, {Anderson},
  {Annis}, {Bahcall}, {Bakken}, {Barkhouser}, {Bastian}, {Berman}, {Boroski},
  {Bracker}, {Briegel}, {Briggs}, {Brinkmann}, {Brunner}, {Burles}, {Carey},
  {Carr}, {Castander}, {Chen}, {Colestock}, {Connolly}, {Crocker}, {Csabai},
  {Czarapata}, {Davis}, {Doi}, {Dombeck}, {Eisenstein}, {Ellman}, {Elms},
  {Evans}, {Fan}, {Federwitz}, {Fiscelli}, {Friedman}, {Frieman}, {Fukugita},
  {Gillespie}, {Gunn}, {Gurbani}, {de Haas}, {Haldeman}, {Harris}, {Hayes},
  {Heckman}, {Hennessy}, {Hindsley}, {Holm}, {Holmgren}, {Huang}, {Hull},
  {Husby}, {Ichikawa}, {Ichikawa}, {Ivezi{\'c}}, {Kent}, {Kim}, {Kinney},
  {Klaene}, {Kleinman}, {Kleinman}, {Knapp}, {Korienek}, {Kron}, {Kunszt},
  {Lamb}, {Lee}, {Leger}, {Limmongkol}, {Lindenmeyer}, {Long}, {Loomis},
  {Loveday}, {Lucinio}, {Lupton}, {MacKinnon}, {Mannery}, {Mantsch}, {Margon},
  {McGehee}, {McKay}, {Meiksin}, {Merelli}, {Monet}, {Munn}, {Narayanan},
  {Nash}, {Neilsen}, {Neswold}, {Newberg}, {Nichol}, {Nicinski}, {Nonino},
  {Okada}, {Okamura}, {Ostriker}, {Owen}, {Pauls}, {Peoples}, {Peterson},
  {Petravick}, {Pier}, {Pope}, {Pordes}, {Prosapio}, {Rechenmacher}, {Quinn},
  {Richards}, {Richmond}, {Rivetta}, {Rockosi}, {Ruthmansdorfer}, {Sandford},
  {Schlegel}, {Schneider}, {Sekiguchi}, {Sergey}, {Shimasaku}, {Siegmund},
  {Smee}, {Smith}, {Snedden}, {Stone}, {Stoughton}, {Strauss}, {Stubbs},
  {SubbaRao}, {Szalay}, {Szapudi}, {Szokoly}, {Thakar}, {Tremonti}, {Tucker},
  {Uomoto}, {Vanden Berk}, {Vogeley}, {Waddell}, {Wang}, {Watanabe},
  {Weinberg}, {Yanny}, \& {Yasuda}}]{yaa+00}
{York}, D.~G., {Adelman}, J., {Anderson}, Jr., J.~E., {et~al.} 2000, \aj, 120,
  1579

\bibitem[{{Zhang} {et~al.}(2008){Zhang}, {Finoguenov}, {B{\"o}hringer},
  {Kneib}, {Smith}, {Kneissl}, {Okabe}, \& {Dahle}}]{zfb+08}
{Zhang}, Y.-Y., {Finoguenov}, A., {B{\"o}hringer}, H., {et~al.} 2008, \aap,
  482, 451

\bibitem[{{Zhao} {et~al.}(2013){Zhao}, {Jia}, {Chen}, {Li}, {Song}, \&
  {Xie}}]{zjc+13}
{Zhao}, H.-H., {Jia}, S.-M., {Chen}, Y., {et~al.} 2013, \apj, 778, 124

\end{thebibliography}

\end{document}